\documentclass[
  pof,
  aip,
  reprint,
  amsmath,amssymb,
  sd,
]{revtex4-1}

\usepackage[T1]{fontenc}
\usepackage[utf8]{inputenc}
\usepackage{amsmath}
\usepackage{amssymb}
\usepackage{enumitem}
\usepackage{graphicx}
\usepackage{refstyle}
\usepackage[
  citecolor=blue,
  colorlinks,
  linkcolor=blue,
  urlcolor=blue,
]{hyperref}
\usepackage[dvipsnames]{xcolor}

\newcommand{\be}{\begin{equation}}
\newcommand{\ee}{\end{equation}} 
\newcommand{\bea}{\begin{eqnarray}}
\newcommand{\eea}{\end{eqnarray}}

\begin{document}

\preprint{AIP/123-QED}

\title{On the energy spectrum of rapidly rotating forced turbulence}
\author{Manohar K. Sharma}
\email{kmanohar@iitk.ac.in}
\affiliation{
 Department of Physics,
  Indian Institute of Technology Kanpur,
  Uttar Pradesh 208016, India
}
\author{Mahendra K. Verma}
\email{mkv@iitk.ac.in}
\affiliation{
 Department of Physics,
  Indian Institute of Technology Kanpur,
  Uttar Pradesh 208016, India
}
\author{Sagar Chakraborty}
\email{sagarc@iitk.ac.in}
\affiliation{
  Department of Physics,
  Indian Institute of Technology Kanpur,
  Uttar Pradesh 208016, India
}
%
\begin{abstract}
In this paper, we investigate the statistical features of the fully developed, forced, rapidly rotating, {turbulent} system using numerical simulations, and model {the} energy {spectrum} that {fits} well with the numerical data. Among the wavenumbers ($k$) larger than the Kolmogorov dissipation wavenumber, the energy is distributed such that the suitably non-dimensionized energy spectrum is ${\bar E}({\bar k})\approx \exp(-0.05{\bar k})$, where overbar denotes appropriate non-dimensionalization. {For the wavenumbers smaller than that of forcing, the energy in a horizontal plane is much more than that along the vertical rotation-axis.} {For} such wavenumbers, we find that the anisotropic energy spectrum, $E(k_\perp,k_\parallel)$ follows the power law scaling, $k_\perp^{-5/2}k_\parallel^{-1/2}$, where `$\perp$' and `$\parallel$' respectively refer to the directions perpendicular and parallel to the rotation axis; this result is in line with the Kuznetsov--Zakharov--Kolmgorov spectrum predicted by the weak inertial-wave turbulence theory for the rotating fluids.
\end{abstract}
\maketitle

\section{Introduction}
Rotating turbulence~\cite{Davidson:book:TurbulenceRotating}---turbulence in rotating fluids---is a commonly occurring phenomenon in the geophysical and the astrophysical flows. It also occurs in the engineering flows like the ones in turbo-machinery and reciprocating engines with swirl and tumble. The large-scale {structures} of the turbulent system {are} affected by rotation {due} to the Coriolis force that acts in the plane perpendicular to the direction of angular velocity.  Intriguingly, the scaling law of the energy spectrum in the inertial range changes with the rotation rate in a way so as to two-dimensionalize the three-dimensional (3D) fluid turbulence. \textcolor{black}{It is well known that}, Kolmogorov~\cite{Kolmogorov:DANS1941Structure,Kolmogorov:DANS1941Dissipation} proposed an inertial range energy spectrum for homogeneous and isotropic 3D hydrodynamics turbulence: $E(k) \sim \epsilon^{2/3} k^{-5/3}$, where $\epsilon$ is the constant energy dissipation rate. The Kolmogorov spectrum is quite universal and observed in plethora of other realistic settings, e.g., shear flows~\cite{Elsinga:PF2016}, viscoelastic fluids~\cite{Valente:PF2016}, buoyancy-driven systems~\cite{Pawar:PF2016,Verma:book:BDF}, and jet flows~\cite{Buchhave:PF2017}.  On the other hand, the energy \textcolor{black}{cascade} for the two-dimensional (2D) turbulence system shows dual behaviour~\cite{Kraichnan:PF1967_2D,Paret:PRL1997,Chen:PRL2006,Vallgren:JFM2011,Boffetta:ARFM2012}: an inverse cascade at large scales with $E(k) \sim k^{-5/3}$ and a forward cascade {of enstrophy} at relatively small scales with $E(k) \sim k^{-3}$.

The energy spectrum becomes more complex in {the} presence of rotation in the turbulent system and there is no unanimity \textcolor{black}{about} the form of the inertial range energy spectrum \textcolor{black}{among the researchers}. The \textcolor{black}{largest} wavenumber upto which the Coriolis force dominants over \textcolor{black}{the nonlinearity}   is specified by \textcolor{black}{the} Zeman wavenumber, $k_{\Omega} \equiv ({\Omega^3}/{\epsilon})^{1/2}$; larger scales corresponding to $k\lesssim k_\Omega$ are affected the most by the rotational effects. \textcolor{black}{Here, $\Omega$ is the magnitude of the angular velocity.}  Be it analytical, numerical, or experimental investigations of either the forced or the decaying rotating turbulent fluid, it is generally found~\cite{ Ibbetson:JFM1975,Hopfinger:JFM1982,Bardina:JFM1985,Jacquin:JFM1990,Zeman:PF1994,Zhou:PF1995,Canuto:PRL1997,Canuto:PF1997_rotation,Yeung:PF1998,Smith:PF1999,Baroud:PRL2002,Galtier:PRE2003,Hattori:PRE2004,Yang:PF2004,Felsot:PRE2005,Smith:JFM2005, Morize:PF2005,Muller:EPL2007,Chakraborty:PRE2007,Chakraborty:EPL2007,Thiele:JFM2009,Mininni:PF2009,Chakraborty:EPJB2010,Sen:PRE2012, Biferale:PRX2016,Baqui:PF2016} that the energy spectrum {for the rotation dominated wavenumbers scales}  as  $k^{-m}$ where $m\in[2,3]$. Apart from these power-law scalings for relatively larger scales, the energy spectrum for \textcolor{black}{the smaller} scales in rotating turbulence is equally enigmatic. In addition to the characteristic energy {spectrum}, the inverse cascade of energy is a prominent feature of rotating turbulence which has been studied experimentally as well as numerically~\cite{Hossain:PF1994,Smith:PRL1996_cross,Smith:PF1999, Morize:PF2005,Mininni:PF2009,Poquet:IOP2013, Yarom:PF2013}. While one observes~\cite{Yeung:PF1998, Campagne:PF2014} forward cascade of energy in the vertical planes (the planes containing the direction of rotation), as far as the horizontal planes are concerned, there usually is  a forward cascade of energy at the small horizontal scales and an inverse energy cascade at the large horizontal scales. The inverse cascade {yields} coherent columnar structures {in the flow}. There are also evidences~\cite{Mininni:PF2009, Bourouiba:JFM2011} of {non local} energy transfers in rotating turbulence.
 
 In {nonrotating} 3D turbulent {fluids}, subsequent to a model proposed by Kraichnan~\cite{Kraichnan:JFM1959} for the kinetic energy spectrum in far dissipation range, Pao~\cite{Pao:PF1965} and Pope~\cite{Pope:book} separately proposed models which are applicable in both the inertial as well as the dissipation ranges. These two model spectra respectively are $E(k) = {\rm K}_{\rm o} \epsilon^{2/3} k^{-5/3} \exp(-\frac{3}{2} \alpha \nu \epsilon^{-1/3} k^{4/3})$, where ${\rm K}_{\rm o}$ is Kolmogorov constant and $\nu$ is the kinematic viscosity; and $E(k) = C \epsilon^{2/3} k^{-5/3} f_{L}(kL) f_{\eta}(k \eta)$, where $C$ is a real constant, $L$ is the integral length scale, $\eta$ is the Kolmogorov length scale, and $f_{L}$ and $f_{\eta}$ are two non-dimensional functions~\cite{Pope:book}. Another widely used typical phenomenological form of  the energy spectrum in far dissipation range of 3D isotropic homogeneous turbulence is given by $E(k) \sim (k/k_\eta)^{\gamma} \exp[-\beta (k/k_\eta)^{n}]$, where $\gamma$, $\beta$, and $n$ are real constants and $k_\eta$ is the Kolmogorov dissipation scale. The values of $\gamma$, $\beta$ and $n$ are \textcolor{black}{often debated} in turbulence community. While $n=1$ (cf. Smith and Reynold\cite{Smith:PF1991} supporting $n=2$) is agreed upon by many researchers~\cite{Sreenivasan:JFM1985,Foias:PF1990,Sanada:PF1992d,Manley:PF1992,Martinez:JPP1997},  there is relatively more disagreement about the {value~\cite{Kraichnan:JFM1959, Kida:PF1987, Domaradzki:PF1992, Chen:PRL1993}} of $\gamma(=3,-1.6,-2, 3.3)$ and {the value~\cite{Kida:PF1987,Chen:PRL1993,Saddoughi:JFM1994}} of $\beta(=4.9, 7.1, 5.2)$. Needless to say, as far as the more complex problem of the forced rotating turbulence is concerned, the issue of such an energy spectrum is \textcolor{black}{even} wider open.
Given the uncertainty in the {two} exponents, \textcolor{black}{the} usage \textcolor{black}{of} this energy spectrum is debatable. In fact, to the best of our knowledge, no one has reported such an energy spectrum for smaller scales that extend far into the dissipation range of the forced rotating fluid turbulence.

Recently, Sharma et al.~\cite{Sharma:PF2018} have investigated the statistical behaviour of \textcolor{black}{the} rapidly rotating decaying turbulence. They have reported that {a} major fraction of the total energy is confined in \textcolor{black}{the} Fourier modes $(\pm 1, 0, 0)$ and $(0, \pm 1, 0)$, which correspond to the largest coherent columnar structure in the flow. These modes \textcolor{black}{contains} approximately \textcolor{black}{$85$} percent of total energy of the system. Further, they have also proposed a model for the energy spectrum: $E(k) \sim C \epsilon^{2/3}_{\omega} k^{-3} \exp[-C(k/k_{d})^2]$ in the region $k \geq 2 k_{\eta}$. Here $k_d \equiv \epsilon^{1/6}/\sqrt \nu $, $\epsilon_{\omega}$ is the enstrophy dissipation rate, $C$ is a constant, and $k_{\eta}$ is the Kolmogorov dissipation wavenumber. There is no reason to believe \emph{a priori} that these results should hold good for the forced rotating turbulence system. In fact, the introduction of forcing at an intermediate scale---corresponding wavenumber being $k_f$---breaks the kinetic energy spectrum into two disjoint region: larger scales $k<k_f$ and smaller scales $k>k_f$.  Contrary to what has been observed in the decaying case, the energy spectrum in the forced case shows power law scaling close to $E(k_{\perp},k_\parallel) \sim k_{\perp}^{-5/2}k^{-1/2}_{\parallel}$ in highly anisotropic large scales. (The subscripts `$\perp$' and `$\parallel$' respectively refer to the directions perpendicular and parallel to the rotation axis.) In the rather isotropic far dissipation range, the energy spectrum becomes exponential function of `$-k$'. Additionally, although we have observed that $(\pm 1, 0, 0)$ and $(0, \pm 1, 0)$ still contain {a major fraction of the total energy}, the energy content in \textcolor{black}{the} intermediate and \textcolor{black}{the} smaller scales \textcolor{black}{is} much more than that for the decaying rotating turbulent; as a result, one observes rather {diffused} columnar structures in case of forced rotating turbulence. The main \textcolor{black}{aim} of our paper is to elaborate on these interesting results and to contrast the forced \textcolor{black}{turbulence} with the decaying \textcolor{black}{one}.

However, before we start presenting the results in a systematic manner, let us clearly describe {our} model system.
 %
\section{The model system} 
\label{Sec:2}
The  governing equation of a forced incompressible fluid flow in \textcolor{black}{the} rotating reference frame is:
\begin{eqnarray}
\frac{\partial \mathbf{u}}{\partial t} + (\mathbf{u} \cdot \nabla) \mathbf{u} & = & -{\boldsymbol{\nabla} p} - 2 {\boldsymbol{\Omega}} \times {\mathbf{u}} + \nu \boldsymbol{\nabla}^2 {\mathbf{u}} + \mathbf{f}, \label{eq:u_dim} \\
\boldsymbol{\nabla} \cdot \mathbf{u} & = & 0 \label{eq:inc_dim}, 
\end{eqnarray}
where $\mathbf{u}$ is the velocity field, $\boldsymbol{\Omega} = \Omega \hat{z}$ is the angular velocity of the rotating frame, $p$ is the pressure field \textcolor{black}{which includes {contributions from}} centrifugal acceleration, $\nu$ is the kinematic viscosity, $-2 \boldsymbol{\Omega} \times \mathbf{u}$ is the Coriolis acceleration, and $\mathbf{f}$ is the {force} field.

We have simulated these equations in \textcolor{black}{a} {cube} of size $(2\pi)^3$ with periodic boundary condition on all the sides using pseudo-spectral code, Tarang~\cite{Verma:Pramana2013tarang,Chatterjee:JPDC2018}. We have used fourth-order Runge-Kutta method for time stepping, and Courant-Friedrich-Lewy (CFL) condition to optimize the time stepping $(\Delta t)$ {and} $2/3$ rule for dealiasing. Our simulations are with  grid-resolutions of $512^3$ and $1024^3$, {and a} constant rotation rate $\Omega = 32$ which {corresponds} to Rossby number (${\rm Ro}$) of the order of $10^{-3}$. One may recall that ${\rm Ro}\equiv {U}/{\Omega L}$, {where} $U$ and $L$ respectively being the large scale velocity and the corresponding length scale, is the ratio of the magnitudes of $(\mathbf{u} \cdot \nabla) \mathbf{u}$ and the Coriolis acceleration.  We have used our forcing scheme in such a way that it supplies constant energy and \textcolor{black}{no} kinetic helicity in the flow. In \textcolor{black}{the} Fourier space the forcing function $(\mathbf{\hat{f}}(\mathbf{k}) )$ is taken as,
\begin{eqnarray}
\mathbf{\hat{f}}(\mathbf{k}) =\frac{\epsilon_{in} \hat{\mathbf{u}}(\mathbf{k}) }{n_f \left[\hat{\mathbf{u}}(\mathbf{k}) \hat{\mathbf{u}}^*(\mathbf{k})\right]},
\end{eqnarray} 
where $n_f$ is the total number of modes at which the forcing is active, and $\epsilon_{in}$ is the energy input rate. The phase of $\mathbf{\hat{f}}(\mathbf{k}) $ is chosen randomly at every time step.
Other relevant parameters of our simulations are tabulated in Table~\ref{table:parameters}.

\setlength{\tabcolsep}{5pt}
\begin{table}[htbp]
\label{table:parameters}
\begin{ruledtabular}
\caption{Parameters of the simulation: the grid-resolution $N$, the forcing wavenumber band $k_{f}$, rotation rate $\Omega$, kinematic viscosity $\nu$, final eddy turn-over time $t_f$, energy supply rate $\epsilon$, Rossby number $\rm{Ro}$, Reynolds number $\rm{Re}$, Zeman wavenumber $k_{\Omega}$, $k_{\rm max} \eta$, and the Kolmogorov dissipation wavenumber $k_{\eta}$.}
\label{table:parameters}
\begin{minipage}[b]{0.6\linewidth}
\begin{center}
\begin{tabular}{ccc}
$N$               &   $512^3$     & $1024^3$ \\ [0.1mm]  
$k_f$              &   $40-41$     & $ 80-82 $ \\ [0.1mm] 
$\Omega$           &   $32$        & $32$ \\ [0.1mm] 
$\nu$              &   $0.001$     & $0.001$   \\ [0.1mm] 
$t_f$              &   $56$        & $3$ \\ [0.1mm] 
$\epsilon$         &   $0.40$      & $0.80$ \\ [0.1mm] 
$\mathrm{Ro}$      &   $0.007$     & $0.005$ \\ [0.1mm] 
$\mathrm{Re}$      &   $1736$      & $1786$ \\ [0.1mm] 
$k_{\Omega}$       &   $286$       & $202$ \\ [0.1mm] 
$k_{\eta}$         &   $141$       & $168$ \\  [0.1mm]
$k_{\rm max} \eta$     &   $4.4$       & $3.0$ \\  [0.1mm]
\end{tabular}
\end{center}
\end{minipage}
\end{ruledtabular}
\end{table}

We have used \textcolor{black}{the} data \textcolor{black}{corresponding to the} time frame \textcolor{black}{$t = 45$} of $512^3$ grid resolution as an initial condition for $1024^3$ grid. The system evolve {up to} $t = 3$ eddy turn-over time for $1024^3$ grid. {We find that} $k_{\rm max} \eta > 1.5$ for the simulation, where $\eta$ is the Kolmogorov's length scale, which shows our simulation is well resolved. We have also ensured that {the} data is collected from the {steady} state. Additionally, by \textcolor{black}{working with} the two grid-resolutions, the results presented in this paper have been shown to be {grid} independent.
\begin{figure}[htbp]
\begin{center}
\includegraphics[scale=1]{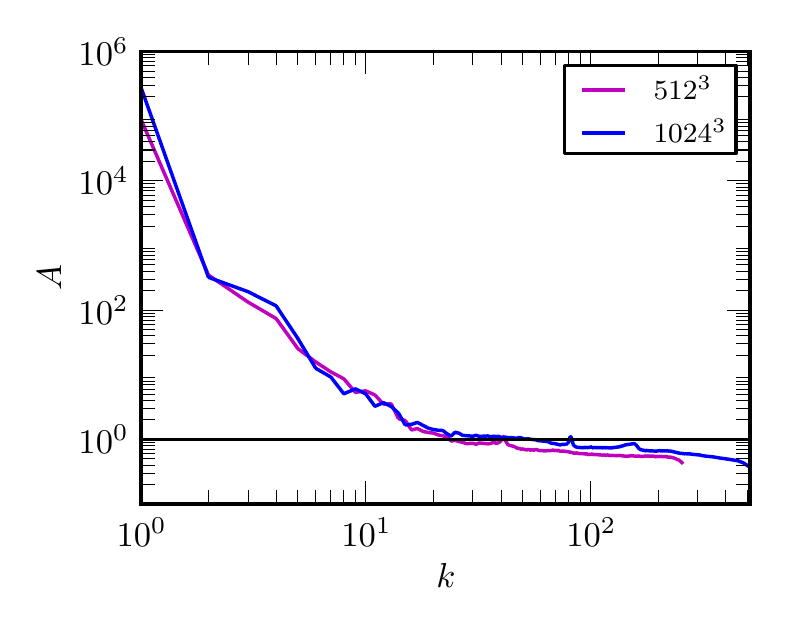}
\caption{ Anisotropy of the forced rotating turbulent system vs. wavenumber at $t = t_f$ for grid simulation of $512^3$ grid resolution (magenta) and $1024^3$ grid resolution (blue).}
\label{fig:Anisotropy}
\end{center}
\end{figure}
\begin{figure}[htbp]
\begin{center}
\includegraphics[scale=1.0]{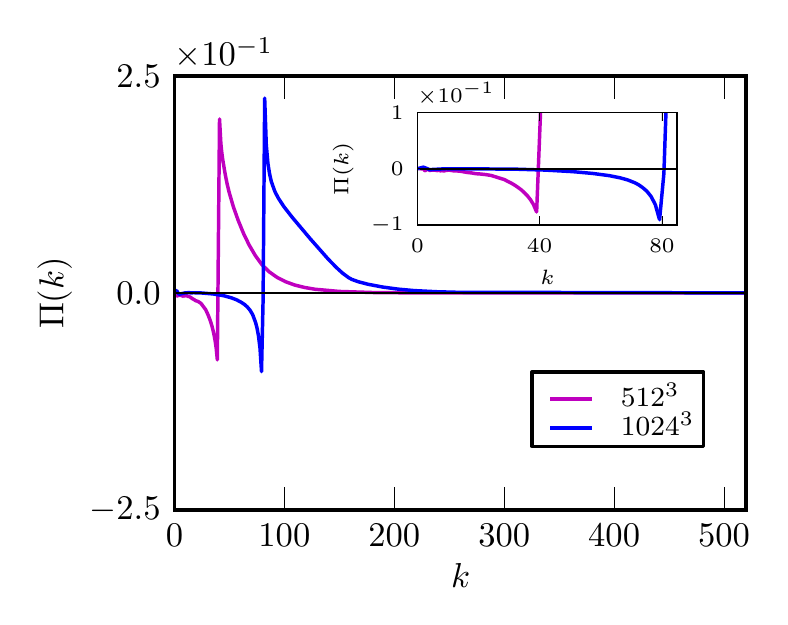}
\caption{ Plot of kinetic energy flux $\Pi(k)$ vs. $k$ of the 3D velocity field at $t = t_f$ for grid resolution $512^3$ (magenta) and $1024^3$ grid resolution (blue). The peaks in the plots \textcolor{black}{indicate} the \textcolor{black}{corresponding} forced wavenumbers $k_{f}$.}
\label{fig:flux}
\end{center}
\end{figure}
\section{Quasi-two-dimensionalization}
Coriolis force affects the perpendicular components of velocity field $\mathbf{u}_{\perp} = u_{x} \hat{\bf x} + u_{y} \hat{\bf y}$, and this force is the primary reason for the quasi-two-dimensionalization (2D) of three-dimensional (3D) turbulent flow. Below we present evidences for {the quasi-two-dimensional nature of the flow}.
\subsection{Strong anisotropy}
We know that, by definition, {the Coriolis force is dominant} at wavenumbers smaller than the Zeman wavenumber. In Fig.~\ref{fig:Anisotropy} we  plot the anisotropy parameter $A\equiv{E_\perp}/{2E_\parallel}$ as a function of wavenumber.  Here, $E_{\perp} \equiv E_{x} + E_{y}$, and $E_{\parallel} \equiv E_{z} $ [with $E_x \equiv \int (u_x^2/2) d{\bf r}$, $E_y \equiv \int(u_y^2/2) d{\bf r}$, and $E_z \equiv \int (u_z^2/2) d{\bf r}$]. It is obvious from the figure that there is strong anisotropy ($A\gg1$), and hence quasi-two-dimensionalization, at larger {scales;} the energy ($E_{||}$) in the parallel (to $\boldsymbol{\Omega}$) component of velocity is much less than the average energy $(E_\bot)$ in one of the perpendicular components. It is of importance to note that at smaller scales, the system is effectively isotropic since $A\sim1$, and consequently the flow can be seen as a 3D turbulent flow where the effect of rotation is negligible.

\subsection{Inverse energy cascade}
The energy evolution equation \textcolor{black}{may be written as},
\begin{eqnarray}
\frac{\partial E(k,t)}{\partial t} &=& T(k,t) - 2 \nu k^2 E(k,t) + F(k,t), \label{eq:energ_evol}
\end{eqnarray}
where $T(k,t)$ is the energy transfer to the wavenumber shell $k$ due to nonlinearity, $-2 \nu k^2 E(k,t)$ is the energy dissipation spectrum, and $F(k,t)$ is the energy injected in the system by forcing. Away from the forcing scales, in steady state where $\frac{\partial E(k,t)}{\partial t} \approx 0$, Eq.~(\ref{eq:energ_evol}) becomes, 
\begin{eqnarray}
\frac{d \Pi(k)}{d k}\equiv -T(k) = -2 \nu k^2 E(k). \label{eq:pi_dissi}
\end{eqnarray}
$\Pi(k)$ is the kinetic energy flux, defined as the net transfer of energy out of the sphere of radius $k$ by the modes inside the sphere. We can compute the energy flux of the system using the following formula~\cite{Dar:PD2001,Verma:PR2004}: 
\begin{eqnarray}
\Pi(k_{*}) = \sum_{k >  k_{*}} \sum_{p \leq k_{*}} S(\mathbf{k} | \mathbf{p} | \mathbf{q}) \label{eq:3D_flux},
\end{eqnarray}
where
\begin{equation}
 S(\mathbf{k}| \mathbf{p}| \mathbf{q}) \equiv \mathrm{Im}[ ({\mathbf{k} \cdot  \mathbf{u}(\mathbf{q})}) ({\mathbf{u}(\mathbf{p}) \cdot  \mathbf{u}^{*}(\mathbf{k})})],\label{eq:Skpq}
\end{equation}
is the mode-to-mode energy transfer rate from the mode $\mathbf{p}$ to mode $\mathbf{k}$ with mode $\mathbf{q}$ as a mediator in a triad ($\mathbf{k}, \mathbf{p}, \mathbf{q}$; $\mathbf{k} = \mathbf{p} + \mathbf{q}$). The resulting figure, Fig.~\ref{fig:flux}, explicitly depicts that {at} the scales larger than the scale at which forcing is active, there is inverse cascade of energy that is reminiscent of similar inverse cascade in \textcolor{black}{the} two-dimensional turbulence. As a result of this inverse cascade of \textcolor{black}{the} kinetic energy, there would be condensation of kinetic energy at lower wavenumber. This is also seen in the decaying rotating turbulent system~\cite{Sharma:PF2018}. This condensation is naturally expected to give rise to \textcolor{black}{the} large scale coherent structures.
\subsection{Large scale coherent structures}

\begin{figure}[htbp]
\begin{center}
\includegraphics[scale=1.0]{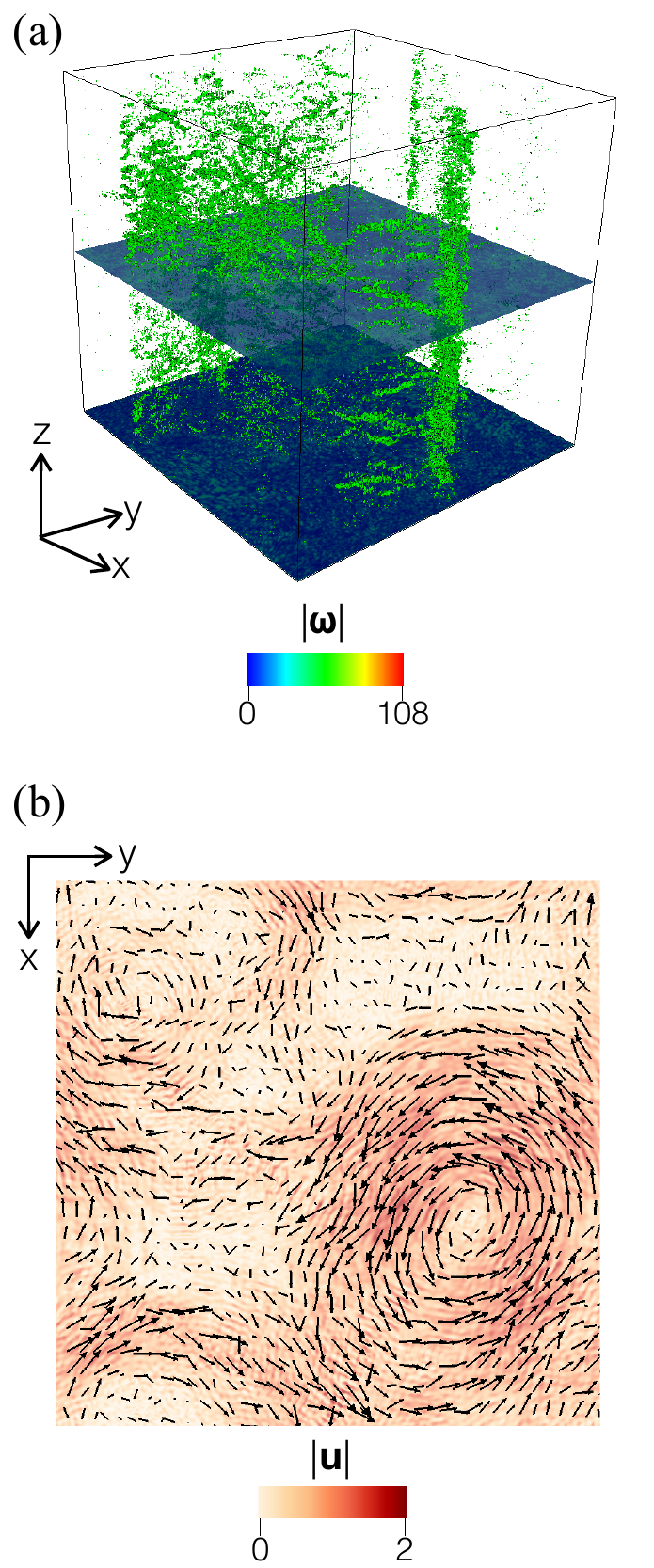}
\caption{Plots for (a) {iso-surfaces} of {various magnitudes} of the 3D vorticity field, and (b) the density plot for the cross-section of 2D velocity field taken at $z = \pi$. Both the plots use the $1024^3$-grid simulations data.}
\label{fig:isosurface}
\end{center}
\end{figure}
There is a {large body of works\cite{Bartello:JFM1994,Godeferd:JFM1999,Smith:PF1999,Morize:PF2005,Staplehurst:JFM2008,Moisy:JFM2011,Gallet:PF2014,Ranjan:JFM2014,Sharma:PF2018} on the formation of \textcolor{black}{the} large scale structures, which is known to depend on the strength of rotation rate: at a very high rotation rate, the flow profile becomes quasi-two-dimensionalized {with} sharp coherent vortical {columns}, while at a low rotation rate, the columnar vortices are understandably disorganized. Figure~\ref{fig:isosurface}(a) exhibits {isosurfaces of constant $|\boldsymbol{\omega}|$ at $t = t_f$, where  $\boldsymbol{\omega} = \nabla \times \mathbf{u}$ is the vorticity field}. {In Fig.~\ref{fig:isosurface}(b)} we {plot} the 2D velocity {($\mathbf{u} = u_x \hat{\bf x} + u_y \hat{\bf y} $)} field for \textcolor{black}{the} horizontal cross-section taken at $z = \pi$ of the flow profile. {The figure shows vortical columns}. We observe {that} the columnar \textcolor{black}{vortices} are disorganized and not very sharp in appearance. Compared to the decaying case\cite{Sharma:PF2018}, in 3D forced rotating turbulence flow, the velocity in the direction of rotation is no longer constant, and the stretching and tilting of vortices occurs. Also, in the plane perpendicular to the direction of rotation, incompressible condition ($\boldsymbol{\nabla} \cdot \mathbf{u}_{\perp} \neq 0$) {is} not satisfied and the two-dimensionalization is a relatively hindered than what happens in the decaying case. Additionally, the forcing {randomizes the flow,} and it starts affecting the velocity field and obstructs the alignment of vorticity field along the direction of rotation. \textcolor{black}{This} obstruction in alignment of velocity fields affects the structure formation in the forced rotating turbulent system.  

In order to quantitatively understand the disorganization of the columnar structures in forced rotating turbulence, we study the energy content of the Fourier modes. Table~\ref{table:total_energy} tabulates the energy {contents} of {the} most energetic Fourier modes for the forced rotating case at time frame $t = 56$ for the grid resolution of $512^3$ and at $t = 3$ for $1024^3$ grid. We do not list the energy of $-\mathbf{k}$ modes because $\mathbf{u}(\mathbf{-k})$ = $\mathbf{u}^{*}(\mathbf{k})$. The fraction of total energy contained in $18 \times 2 = 36$ Fourier modes is a significant amount of \textcolor{black}{the} total energy spread over $512^3$ {and} $1024^3$ modes. The Fourier modes $(k_x, k_y, k_z) = (1, 0, 0)$ and $(0, 1, 0)$ are the most dominant modes of the system. These modes {contain} $25$ {to} $30$ percent of total energy of the system. The situation is different in case of decaying rotating turbulence, {where} modes {$(\pm 1,0,0)$ and $(0, \pm 1, 0)$} contain arround {$90$ }percent of total energy ~\cite{Sharma:PF2018}.  It may be remarked that in case of decaying turbulence these modes are not strong initially but later in time they become dominant. In comparison, however, there is no significant {temporal variation} in the kinetic energy distribution for {the} forced rotating turbulence, as can be gathered from Table~\ref{table:total_energy}. The important thing to note is that, in case of \textcolor{black}{the} forced rotating turbulent system, more energy is uniformly distributed in the other modes as should be expected for \textcolor{black}{the} relatively less coherent columns.
\setlength{\tabcolsep}{5pt}
\begin{table}[htbp]
\begin{ruledtabular}
\caption{Percentage energy distribution {among} dominant modes {at} $t=56$ and $t = 3$ for the grid resolution of $512^3$ and $1024^3$ at the rotation rate $\Omega=32$ respectively. In table $E_{\mathrm{mode}}$ is defined as $E_{\mathrm{mode}} = |u(\mathbf{k})|^2/2$, and $E$ is the total energy of the system at $t = 3, 56$.} 
\label{table:total_energy}
\begin{minipage}[b]{0.7\linewidth}
\begin{tabular}{l  c c  }
\phantom{xx}Mode &   $E_{\mathrm{mode}}/E$  (\%)   & $E_{\mathrm{mode}}/E$   (\%) \\
($k_x$, $k_y$, $k_z$)  & $t = 56$      &  $t = 3.0$   \\ [0.1mm] 
                       &  $ N = 512^3$        &  $N = 1024^3$  \\ [0.1mm]
\hline 
$(1,0,0)$              & $14.17$     & $15.02 $ \\ [0.1mm]
$(0,1,0)$              & $11.31$     & $13.97$  \\ [0.1mm]
$(1,1,0)$              & $1.62$      & $2.82 $  \\ [0.1mm]
$(-1,1,0)$             & $1.43$      & $0.45 $  \\ [0.1mm]
$(1,2,0)$              & $0.92$      & $1.77$  \\ [0.1mm]
$(2,-1,0)$             & $0.51$      & $0.40$  \\ [0.1mm]
$(-2,1,0)$             & $0.51$      & $0.40$  \\ [0.1mm]
$(2,2,0)$              & $0.34$      & $0.04$  \\ [0.1mm]
$(1,-2,0)$             & $0.31$      & $0.19$ \\ [0.1mm]
$(-3,1,0)$             & $0.31$      & $0.22$  \\ [0.1mm]
$(2,-2,0)$             & $0.21$      & $0.12$  \\ [0.1mm]
$(3,1,0)$              & $0.13$      & $0.06$  \\ [0.1mm]
$(3,-2,0)$             & $0.13$      & $0.06$  \\ [0.1mm]
$(2,1,0)$              & $0.13$      & $0.23$  \\[0.1mm] 
$(-1,-3,0)$            & $0.07$      & $0.15$  \\ [0.1mm]
$(-3,-2,0)$            & $0.05$      & $0.44$  \\ [0.1mm]
$(-2,-3,0)$            & $0.02$      & $0.13$  \\ [0.1mm]
$(-3,0,0)$             & $0.02$      & $0.06$  \\ [0.1mm] 
\hline\hline
Total \%:              & $32.19$     & $ 36.53 $\\ [0.1mm]
\end{tabular}
\end{minipage}
\end{ruledtabular}
\end{table}

\begin{figure}[htbp]
\begin{center}
\includegraphics[scale=1.0]{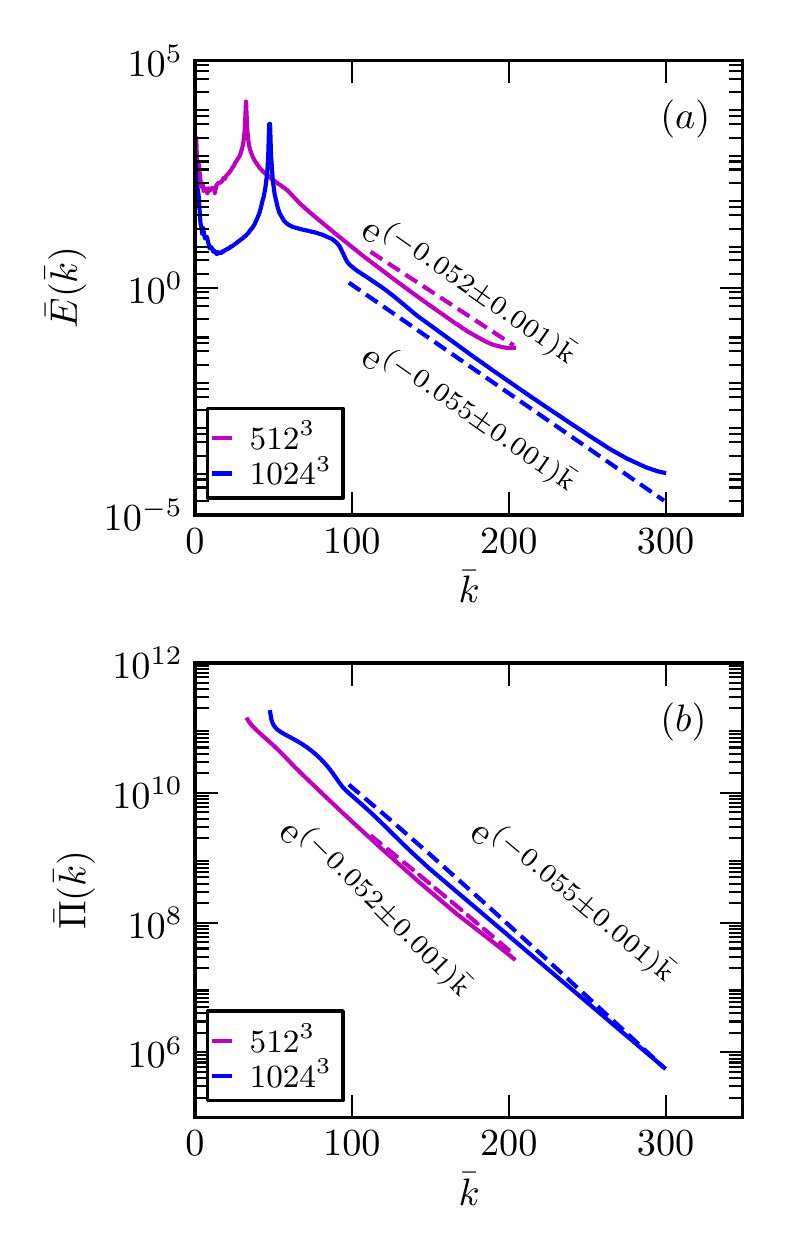}
\caption{ Plots of (a) the non-dimensionalized energy spectra, $\bar{E}(\bar{k})$,  and (b) the non-dimensionalized energy flux,  $\bar{\Pi}(\bar{k})$ at $t=t_f$, as generated by the simulations done with grid resolutions of $512^3$  (magenta) and $1024^3$ (blue) for the forced rotating turbulent fluid. The dotted lines are the fits generated in accordance with the energy spectra and the energy flux given in Eqs.~(\ref{eq:model_energy})-(\ref{eq:model_flux}).}
\label{fig:model_exponential}
\end{center}
\end{figure}
\begin{figure*}[htbp]
\begin{center}
\includegraphics[scale=1.0]{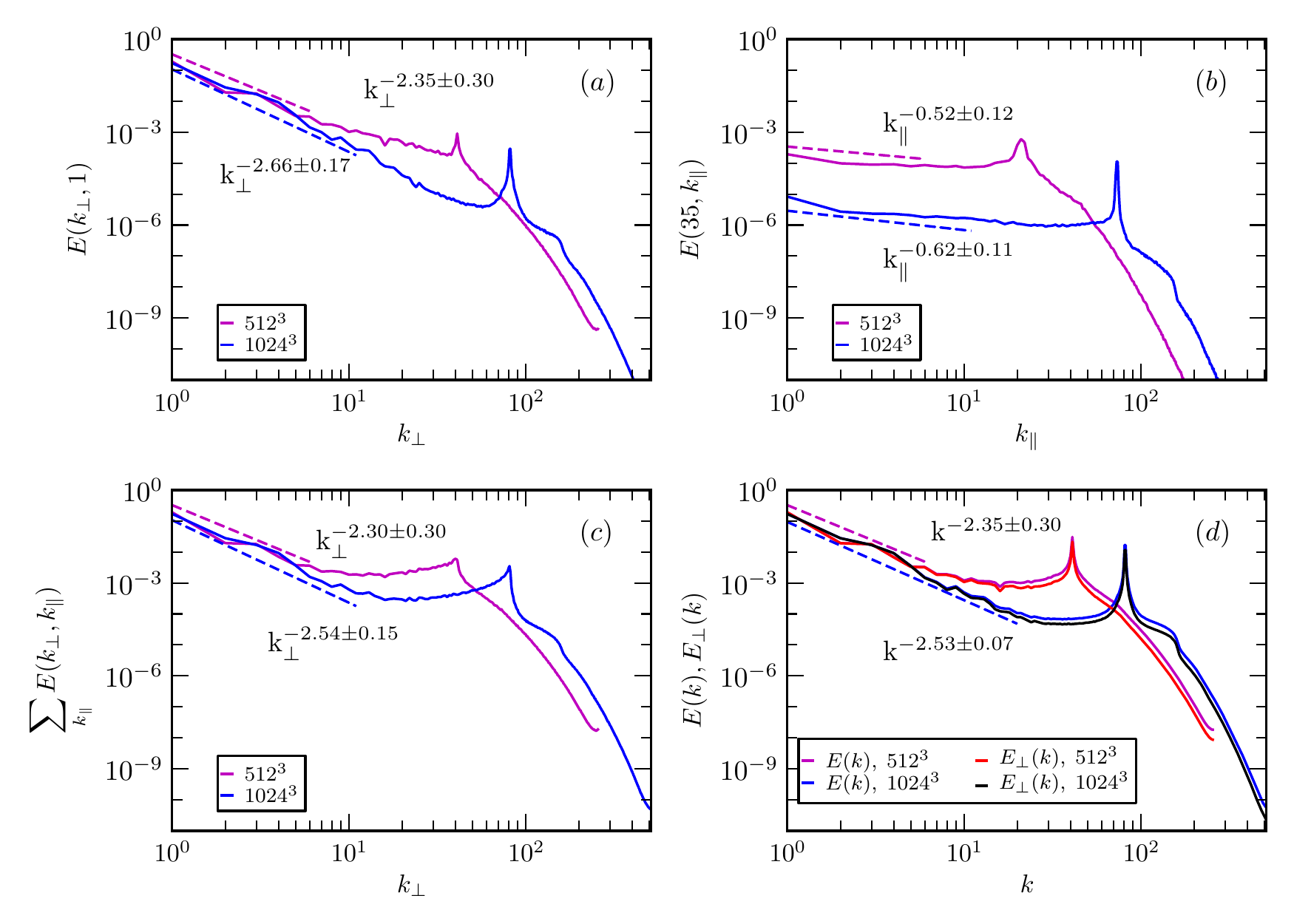}
\caption{Plots of the (aniotropic) energy spectrum, ${E}({k_\perp,k_\parallel})$ at $t=t_f$, as a function of (a) $k_{\perp}$ for $k_\parallel=1$, (b) $k_{\parallel}$ for $k_\perp=35$, and (c) $k_\perp$ after the spectrum is summed over all $k_{\parallel}$. The dashed lines are the fitting lines; we have appropriately used the wavenumber range $1\le k_{\perp}, k_{\parallel}\le 6$ for the $512^3$ grid and $1\le k_{\perp}, k_{\parallel}\le 11$ for the $1024^3$ grid as the fitting ranges. As in the other plots, we have used magenta for $512^3$ grid-resolution data and blue for $1024^3$ grid-resolution data. Subplot (d) exhibits the plots for the (isotropic) energy spectra, $E(k)$ (magenta and blue) and $E(k_{\perp})$ (red and black), as functions of $k$.  Here, we have used additional colours---red and black---to respectively indicate the data corresponding to $512^3$ grid and $1024^3$ grid. The fitting ranges in subplot (d) are $1\le k \le 6$ for $512^3$ grid and $1\le k \le 20$ for $1024^3$ grid.}
\label{fig:scaling_law}
\end{center}
\end{figure*}
\section{The Energy Spectrum}
\label{Sec:3}
The energy spectrum across the full range of wavenumbers is shown in Fig.~\ref{fig:model_exponential}(a) and Fig.~\ref{fig:scaling_law}. We note that about the forced scales (corresponding to the peaks in the plots), the spectrum is too disrupted to show any prominent scaling. We, thus, focus our attention on two different non-overlapping ranges of scales: the scales smaller than the forcing scales and the scales larger than the forcing scales. 

It is obvious that the larger scales, being more affected by the rotation, should behave as if they are quasi-two-dimensionalized. In contrast, the smaller scales should have energy spectrum with an exponential term reminding one of the far dissipation range of the 3D isotropic homogeneous turbulence and the exponential fit found for the rapidly rotating decaying turbulence\cite{Sharma:PF2018}. However, the energy spectra that we report in this paper for both the ranges ($k<k_f$ and $k>k_f$) are---to the best of our knowledge---unreported in any direct numerical simulations done so far. In particular, we find that while the exponential fit in the decaying case goes as $\exp(-{\rm constant}\times k^2)$, in the forced case it goes as $\exp(-{\rm constant}\times k^1)$ that is less steep in the dissipation range $(k>k_\eta,k_f)$. Furthermore, in the anisotropic regime, i.e., in the larger scales $(k<k_f)$, the Kuznetsov--Zakharov--Kolmgorov (KZK) spectrum\cite{Galtier:PRE2003} emerges. In what follows, we model and discuss these spectra in detail as discovered in our numerical simulations.
\subsection{Model spectrum for smaller scales}
\label{sub:model_laminar}
 Most models of the forced rotating turbulence predict Kolmogorov's $k^{-5/3}$ spectrum in the wavenumber range: $k_\Omega, k_f < k < k_\mathrm{DI}$, where $ k_\mathrm{DI}$ is the transition wavenumber (usually much smaller than $ k_\eta$) between the inertial range and the dissipation range. Thus, in order to see this scaling, at the lower end we need to pick the larger one between $ k_\Omega$ and $ k_f$. Unfortunately, in all our simulations, $k_\Omega>k_\eta$ and $k_f$ is very close to $k_\eta$ that is approximately just two to three times more than $k_f$. Because of this, there is no significant range of wavenumbers that could exhibit $k^{-5/3}$ spectrum discernible in a log-log plot. 

Recently, Verma {\em et al.}\cite{Verma:arxiv2017} and Verma\cite{Verma:book:BDF} have shown that for laminar hydrodynamic flows, the steady state energy spectrum and the flux are proportional $\exp(-k/k_\eta)/k$ and  $\exp(-k/k_\eta)$ respectively.  These functions satisfy Eq.~(\ref{eq:pi_dissi}), and also match with numerical simulations of the laminar flows. For our rapidly rotating flow, the energy content in the scales smaller than the forcing scale,  $\sum_{k > k_f} E(k)$, is quite small; hence it can be treated as approximately laminar.   This is due to the strong inverse cascade of energy to the larger scales that retain most of the energy.   Motivated by this observation,  we propose the following form for $E(k)$ in the smaller scales extending into the far dissipation range:
\begin{eqnarray}
E(k) \equiv \sum_{k-1<k'\leq k} \frac{1}{2} |\hat{\mathbf{u}}(\mathbf{k}')|^2= u^{2}_{\mathrm{rms}} f_{L}(k)\frac{1}{k} \exp(-\alpha k/k_d).
\label{eq:model_ene}\nonumber\\ \label{eq:far_energy_spectrum}
\end{eqnarray}
From the form of this $E(k)$, it may be noted that Eq.~(\ref{eq:pi_dissi}) contains the effects of weak nonlinearity of the flow. The weak energy flux $\Pi(k)$ is the result of this nonlinearity. 

The above $E(k)$ has only {one} exponent, viz., $\alpha$, that needs to be determined. Here $u_{\mathrm{rms}}$ is the rms velocity of the high-pass filtered flow with all the component wave vectors greater than $k_\eta$. Note that we have introduced a new wavenumber $k_d$ that is taken to be \textcolor{black}{$\sqrt{Re}/L$}, but now $Re$ is calculated with $U$ replaced by $u_{\mathrm {rms}}$. In this context, recall that $k_{\eta} \equiv \left({\epsilon}/{\nu^3}\right)^{1/4}\sim\sqrt{Re}/L$, where
\begin{eqnarray}
\epsilon &=& \int^{\infty}_{0} 2 \nu k^2 E(k)dk=2 \nu u^2_{\mathrm{rms}} k_d^2 I, \label{eq:const_A}
\end{eqnarray}
with $I \equiv \int_{0}^{\infty} \bar{k} f_{L}(\bar{k}) \exp(-\alpha \bar{k}) d\bar{k}$ and $\bar{k}\equiv {k}/{k_{d}}$. Thus, $k_d$ based on the high-pass filtered flow is analogous to $k_\eta$ that is based on the unfiltered flow field.

Using $\bar{k}$, $\bar{E}(\bar{k}) \equiv {E(k) k }/{u^2_{\mathrm{rms}}}$, and $\bar{\Pi}(\bar{k})  \equiv {\Pi(k)}/{\epsilon}$ as non-dimensionalized quantities, the relationship [Eq.~(\ref{eq:pi_dissi})] between the kinetic energy and the energy flux in the non-dimensionalized form becomes:
\begin{eqnarray}
\frac{d \bar{\Pi}(\bar{k})}{d\bar{k}} &=& -\frac{\bar{k}}{I} \bar{E}(\bar{k})\label{eq:dpi_dk}.
\end{eqnarray}
In the dissipation range $k\gg k_f$, where $k_f$ is the forcing wavenumber and $f_{L}(\bar{k}) $ is unity,  the kinetic energy and the energy flux are respectively as follows:
\begin{eqnarray}
\bar{E}(\bar{k}) &=& \exp(-\alpha \bar{k}), \label{eq:model_energy}\\ 
\bar{\Pi}(\bar{k}) &=& \frac{1}{I} \left( \frac{1}{\alpha^2} + \frac{\bar{k}}{\alpha} \right) \exp(-\alpha \bar{k}). \label{eq:model_flux}
\end{eqnarray}
It may be easily checked that the above two expressions satisfy the non-dimensionized equation, {Eq.~(\ref{eq:dpi_dk})}.

In Fig.~\ref{fig:model_exponential}(a)-(b), we {plot} $\bar{E}(\bar{k})$ vs. $\bar{k}$ and $\bar{\Pi}(\bar{k})$ vs. $\bar{k}$ respectively in semi-log scale at time frames $t = 56$ (magenta) for grid resolution of $512^3$ and at $t = 3$ (blue) for $1024^3$ grid. We observe that the predicted self-consistent model [Eqs.~(\ref{eq:model_energy})-(\ref{eq:model_flux})] of the energy spectrum and the energy flux stands validated by our numerical data in the wavenumber range $k \in [141,256]$ for $512^3$ grid resolution, and $k \in [168,512]$ for $1024^3$ grid resolution. The {values} of $\alpha$ {are} $0.052 \pm 0.001$ for $512^3$ grid resolution, and $0.055 \pm 0.001$ for $1024^3$ grid. We are getting approximately same value of $\alpha$ for both the grid resolutions, thus, signifying that the parameter is robust and resolution-independent. The values of ${\rm Re}$ based on the aforementioned wavenumber ranges are $62$ for $512^3$,  and $115$ for $1024^3$ {resolutions, which} are one order less than the global $\textrm{Re}$ {( $1736$ and $1786$ respectively for the grid resolutions $512^3$ and $1024^3$)}. Thus, there is reduction in strength of turbulence in the smaller scales, {that is the reason why} any power law scaling is hard to observe {here}.
\subsection{Model spectrum near larger scales}
Although one can extract scaling exponents from the $E(k)$~vs.~$k$ plot to characterise the system, strictly speaking, in the anisotropic regime, the energy spectrum should not be modelled as a function of $k$. It is more sensible that the energy spectrum be \textcolor{black}{explicitly dependent} on both ${\bf k}_{\perp}$ (two dimentional vector perpendicular to the rotation axis) and $k_{\parallel}$ (wavenumber corresponding to the direction parallel to the rotation axis). Thus, we define such a spectrum as,

\begin{eqnarray}
E({k_{\perp}}, k_{\parallel}) &=& \sum_{\substack{k_{\perp}-1<k'_{\perp}\leq k_{\perp}, \\ k_{\parallel} - 1 < k'_{\parallel} \le k_{\parallel} }} \frac{1}{2} |\hat{\mathbf{u}}({k'_{\perp}}, k'_{\parallel})|^2. \label{eq:energy_perp_parallel} 
\end{eqnarray}
Note that, under the assumption of axisymmetry, we have used the magnitude of ${\bf k}_{\perp}$ in the argument of the energy spectrum. 

The energy spectrum of the system is calculated by using Eq.~(\ref{eq:energy_perp_parallel}) and plotted in Fig.~\ref{fig:scaling_law}. In Fig.~\ref{fig:scaling_law}(a)-(b), it is remarkable to note that the energy spectrum observed from our numerical simulation is {quite} close to the Kuznetsov--Zakharov--Kolmgorov (KZK) spectrum---$E(k) \sim k_{\perp}^{-5/2} k_{\parallel}^{-1/2}$---within errorbars. The strong rotation supplies a small parameter, $\textrm{Ro}$, in the framework of weak turbulence. Using this it \text{has} shown\cite{Galtier:PRE2003} that, in the anisotropic limit $k_\perp\gg k_{\parallel}$, structures elongated along the rotation axis are brought forth through dominating local interactions. Further, the KZK spectrum comes out as an exact solution. Although the aforementioned weak turbulence analysis has been done for decaying rotating turbulent fluid, we emphasize that this spectrum appears in our numerical simulations exactly where the system is strongly anisotropic and shows elongated structures, thereby somewhat satisfying the requirements for the appearance of the spectrum. It may be noted, in this context, that the aforementioned plots are for the case: $k_\perp\gg k_{\parallel}$, just what the weak turbulence analysis requires. We also note that the shift of the peaks away from $k_f$ in Fig.~\ref{fig:scaling_law}(b) is because the abscissa is $k_\parallel$ and not $k$, and thus, the identity $k^2 = k_{\perp}^2 + k_{\parallel}^2$ with $k=k_f$ and $k_{\perp}=35$ locates the peaks at $k_{\parallel}\approx 21$ and $k_\parallel\approx 74$ respectively for $512^3$ and $1024^3$ grid resolutions.

Additionally,  in Fig.~\ref{fig:scaling_law}(c), we note that the scaling exponent of $k_\perp$ is unchanged even when the spectrum is summed over all $k_\parallel$. Further, Fig.~\ref{fig:scaling_law}(d) showcases the fact that this scaling exponent ($\sim-5/2$) is very robust; in the range $A\gg1$, even the isotropic energy spectrum, $E(k)$, scales as $\sim k^{-5/2}$ and so does $E_\perp(k)$ simply because the share of energy contained in the plane perpendicular to the rotation axis is large. We remark, however, the scaling exponent of $k_\parallel$ is not very robust {and it varies with the choice of $k_{\perp}$ range} (see Appendix~\ref{ap:aniso_limit}). Nevertheless, {we believe that ours is the first numerical demonstration of KZK spectra in forced rotating turbulence.}


\section*{Discussions and Conclusion}
\label{sec:conclusion}

We have performed \textcolor{black}{direct} numerical simulations of a 3D turbulent fluid forced at intermediate scales. {The fluid is rotating} with a high rotation rate corresponding to $\textrm{Ro}\sim10^{-3}$. Such a system is known to mimic some of the features of the 2D turbulence. Recall that in the fully developed 2D fluid turbulence---extensively investigated\cite{Borue:PRL1993,Smith:PRL1993,Paret:PRL1997,Boffetta:EPL2007,Rutgers:PRL1998,Gotoh:PRE1998,Schorghofer:PRE2000,Lindborg:PF2000,Ishihara:PF2001,Chen:PRL2003,Chen:PRL2006} numerically as well as experimentally---strong vortical structures appear in the system due to inverse cascade of kinetic energy from the forcing scale to the scale corresponding to the box-size. Further, it has been shown that the kinetic energy spectrum for $k<k_f$ is given by $E(k)\sim k^{-5/3}$, whereas\cite{Borue:PRL1994, Tran:PRE2004, chertkov:PRL2007, Fischer:PF2009} for $k>k_f$, $E(k) \sim k^{-3}$. 

In the forced rotating flow, we observed that the system becomes highly anisotropic at larger scales; coherent columnar structures are formed but they are relatively more diffused in appearance compared to what is seen in the case of decaying rotating turbulence. We have also found dual cascade regions---forward and reverse cascades---for both the energy and the enstrophy (calculated on a 2D horizontal plane; see Appendix~\ref{ap:inverse_enstrophy}). While these features are very much reminiscent of the 2D turbulence, the forms of the energy spectrum on the either side of $k_f$ is completely different from what is observed in the 2D turbulence. \textcolor{black}{Thus,} the quasi-two-dimensionalization in the rotating 3D turbulent fluid doesn't actually mean that the spectral properties of the fluid become akin to that possessed by the 2D turbulent fluid.

For the anisotropic regime, we have also investigated the behaviour of the kinetic energy spectrum as a function of $k_{\perp}$ and $k_{\parallel}$, which is convenient to showcase the manifestations of anisotropic behaviour of system. We  observed that the kinetic energy spectrum {varies} as $k_{\perp}^{-2.35 \pm 0.30 }$ in wavenumber range $k_{\perp} \in [1,6]$ for grid resolution of $512^3$, and scales as $k_{\perp}^{-2.66 \pm 0.30 }$ in wavenumber range $k_{\perp} \in [1,11]$ for grid resolution of $1024^3$. The kinetic energy spectrum as a function of $k_{\parallel}$ shows power scalings for fixed $k_{\perp}$, specifically, $E(35, k_{\parallel}) \sim k_{\parallel}^{-0.52 \pm 0.12}$ in wavenumber range $k_{\parallel} \in [1,6]$ for grid resolution of $512^3$ and  $E(35, k_{\parallel}) \sim k_{\parallel}^{-0.62 \pm 0.11}$ in wavenumber range $k_{\parallel} \in [1,11]$ for $1024^3$ grid. Within the error bars, these scalings are in conformity with the KZK-spectrum. As far as the isotropic small scales are concerned, we have proposed a model~[Eq.~(\ref{eq:far_energy_spectrum})] that faithfully captures the behaviour of the kinetic energy spectrum in far dissipation range. Our proposed model is in very good agreement with the numerical results over the wavenumber range $k \in [k_{\eta}, k_{\mathrm{max}}]$.

Our investigation highlights that the statistical features of the decaying rotating turbulence and \textcolor{black}{the} forced rotating turbulence are quite different. This fact suggests that a similar scientific comparison of the decaying versus the forced turbulences under the simultaneous effects of the rotation and the non-zero kinetic helicity is worth pursuing in future. This is important because, in any realistic experiment of the forced rotating turbulence, the experimental set-ups are such that some finite amount of the kinetic helicity would invariably be imparted to the fluid. {Also, the kinetic helicity is significant inside the Earth's outer core, and is similarly important in the other planets and the stars.} 

\section*{Acknowledgements}
 The authors thank Abhishek Kumar for many important suggestions. MKV thanks the Department of Science and Technology, India (INT/RUS/RSF/P-03) and Russian Science Foundation Russia (RSF-16-41-02012) for the Indo-Russian project.  SC gratefully acknowledges financial support from the INSPIRE faculty fellowship $(\text{DST}/\text{INSPIRE}/04/2013/000365)$ awarded by the INSA, India and DST, India. The simulations were performed on the HPC2010 and HPC2013 and Chaos cluster of IIT Kanpur, India. 
 
\appendix
\section{Inverse Enstrophy Cascade}
\label{ap:inverse_enstrophy}
\begin{figure}[htbp]
\begin{center}
\includegraphics[scale=1.0]{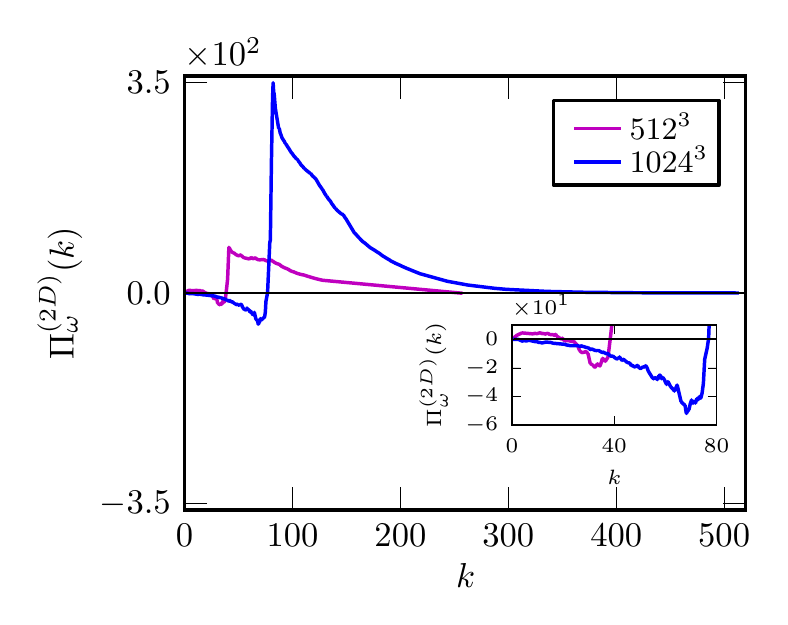}
\caption{ Plot of the 2D enstrophy flux, $\Pi^{(2D)}_{\omega}(k)$, for the 2D velocity field on the horizontal cross-section, $z = \pi$, at $t = t_f$ for the grid-resolutions $512^3$ (magenta)  and $1024^3$ (blue). The inset highlights the inverse cascade of the flux.  The peaks in the plots \textcolor{black}{indicate} the \textcolor{black}{corresponding} forced wavenumbers $k_{f}$.}
\label{fig:enstrophyflux}
\end{center}
\end{figure}
\begin{figure}[htbp]
	\begin{center}
		\includegraphics[scale=1.0]{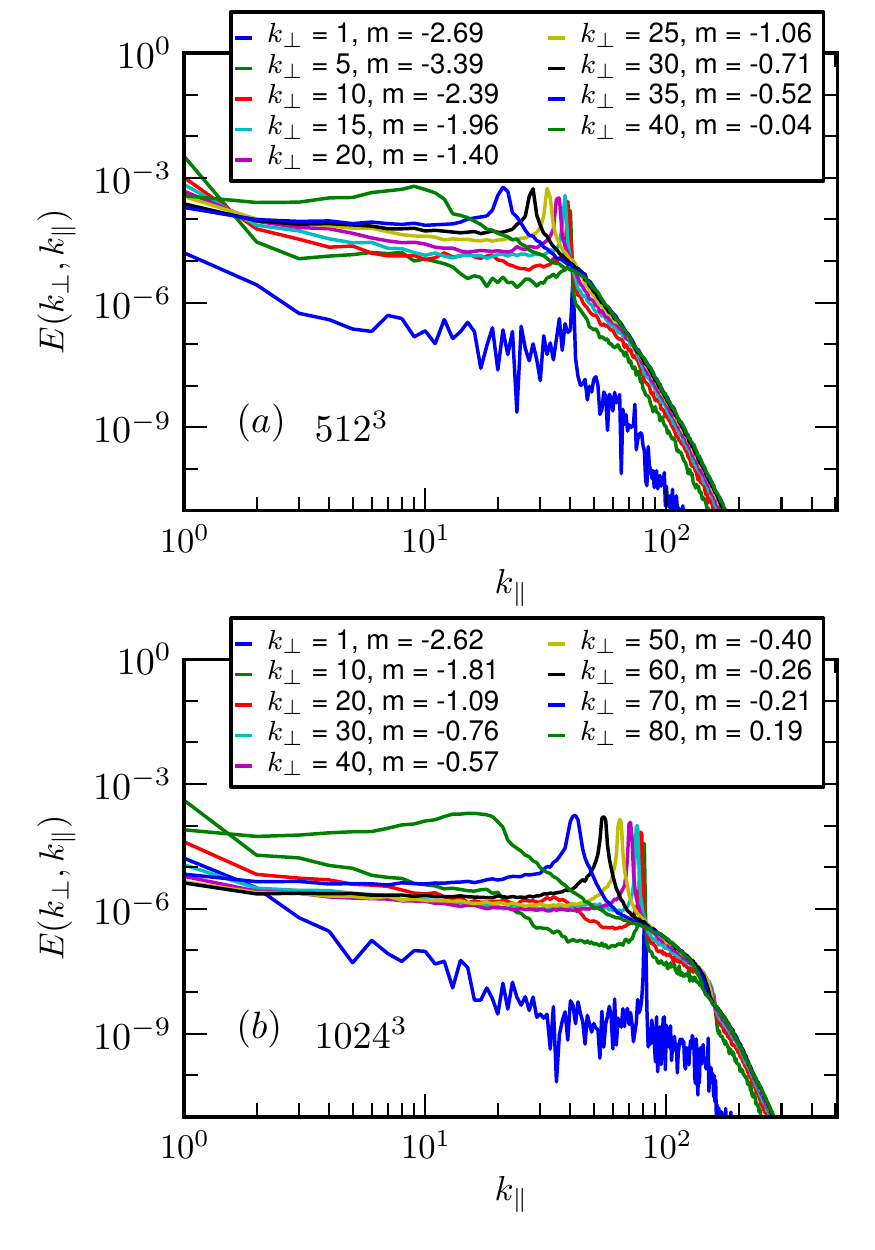}
		\caption{ Plots of the (aniotropic) energy spectrum, ${E}({k_\perp,k_\parallel})$ at $t=t_f$, as a function of $k_{\parallel}$ with different $k_\perp$ for (a) $512^3$ grid resolution and (b) $1024^3$ grid resolution. The fitting exponent ($m$) in $E(k_{\perp}, k_{\parallel}) \sim k_{\parallel}^{-m}$ has been calculated in the wavenumber range $1\le k_{\parallel}\le 6$ for the $512^3$ grid resolution and $1\le k_{\parallel}\le 11$ for the $1024^3$ grid resolution is shown in the legends.}
		\label{fig:aniso_limit}
	\end{center}
\end{figure}
We studied the enstrophy flux of plane perpendicular to the axis of rotation. We compute the enstrophy flux of the $2D$ velocity field $\mathbf{u}_{\perp}$ at the plane using \textcolor{black}{the formula}:
\begin{eqnarray}
\Pi_{\omega}^{(2D)}(k_{*}) &=& \sum_{k>k_{*}} \sum_{p \leq k_{*}} S^{\omega \omega} (\mathbf{k} | \mathbf{p} | \mathbf{q}). \label{eq:enstrophy_flux_2D} 
\end{eqnarray}
\textcolor{black}{Here},
\begin{eqnarray}
S^{\omega \omega} (\mathbf{k} | \mathbf{p} | \mathbf{q}) &=& \mathrm{Im}\left[\left(\mathbf{k} \cdot \mathbf{u}_{\perp}(\mathbf{q})\right)\left(\omega_z(\mathbf{p}) \omega_z^* (\mathbf{k})\right)\right],
\end{eqnarray}
with,
\begin{eqnarray}
\omega_z(\mathbf{k}) = \left[i\mathbf{k} \times \mathbf{u}(\mathbf{k})\right]_z,
\end{eqnarray}
represents the enstrophy transfer from mode $\omega_z(\mathbf{p})$ to mode $\omega_z(\mathbf{k})$ with mode $\mathbf{u}(\mathbf{q})$ acting as a mediator. Figure~\ref{fig:enstrophyflux} illustrate \textcolor{black}{the} enstrophy flux of the horizontal cross-section taken at $z = \pi$ at time frame $t = 56$ (magenta) for grid resolution of $512^3$ and at $t = 3$ (blue) for $1024^3$ grid. Arguably~\cite{Fischer:PF2009}, when the vortex {merger} is strong in 2D turbulent flow, the mean enstrophy flux is negative~\cite{Ohkitani:PFA1991, Babiano:JFM2007} for the scales larger than the injection scales. Therefore, the negative enstrophy flux in the scales larger than the injection scale (as exhibited in Fig.~\ref{fig:enstrophyflux}) signifies that our rotating 3D flow is appreciably two-dimensionalized.
\section{Anisotropic Energy Spectrum's Scaling With $k_\parallel$}
\label{ap:aniso_limit}
The KZK spectrum exists in the anisotropic limit, viz., $k_{\perp} \gg k_{\parallel}$, in rapidly rotating turbulence as was shown by Galtier~\cite{Galtier:PRE2003} (and we have observed so in Fig.~\ref{fig:scaling_law}); however, that analysis did not include forcing. Since our investigation is for the forced rotating turbulent fluid, we must keep this added complication in mind while extracting the exponent of $k_\parallel$ in the anisotropic energy spectrum. The scaling of the energy spectrum as a function of $k_{\parallel}$ depends on the choice of $k_{\perp}$ in the anisotropic limit, which is shown in Fig.~\ref{fig:aniso_limit}.

In Fig~\ref{fig:aniso_limit}, we plot the energy spectrum as a function of $k_{\parallel}$ for different $k_{\perp}$ for both the $512^3$ grid resolution and the $1024^3$ grid resolution. In order to extract the exponent, we have to not only decide what range of $k_\parallel$ to use but also what value of $k_\perp$ to choose. Although, one would like to choose a large enough range while maintaining $k_\perp\gg k_\parallel$, we are restricted in our choice because the peak corresponding to the forcing scale restricts us: the upper bound of the fitting range should ideally be as much away from the forcing scale as possible but the peak shift towards lower wave number, in line with the identity---$k^2=k_\perp^2+k_\parallel^2$,  as $k_\perp$ increases, thereby making the range smaller. This is at odds with the fact that $k_\perp$ must be much greater than $k_\parallel$ for witnessing the KZK spectrum. Thus, it effectively becomes a matter of systematic investigation where one should work with all possible combinations of the $k_\parallel$-ranges and the value of $k_\perp$. What we can definitely conclude from Fig.~\ref{fig:aniso_limit} is that for $512^3$ grid resolution, the range $1\le k_\parallel\le6$ and $k_\perp=35$ make an ideal combination which leads to the observation of the KZK-spectrum in the anisotropic limit in the forced rotating turbulence. Similar conclusion can be made for the $1024^3$ grid resolution with the range $1\le k_\parallel\le11$ and $k_\perp=35$.   
%

\bibliography{../bib_mkv/journal,../bib_mkv/book,../bib_mkv/conf,../bib_mkv/preprint}

\begin{thebibliography}{10}%
\makeatletter
\providecommand \@ifxundefined [1]{%
 \ifx #1\undefined \expandafter \@firstoftwo
 \else \expandafter \@secondoftwo
\fi
}%
\providecommand \@ifnum [1]{%
 \ifnum #1\expandafter \@firstoftwo
 \else \expandafter \@secondoftwo
\fi
}%
\providecommand \enquote [1]{``#1''}%
\providecommand \bibnamefont  [1]{#1}%
\providecommand \bibfnamefont [1]{#1}%
\providecommand \citenamefont [1]{#1}%
\providecommand\href[0]{\@sanitize\@href}%
\providecommand\@href[1]{\endgroup\@@startlink{#1}\endgroup\@@href}%
\providecommand\@@href[1]{#1\@@endlink}%
\providecommand \@sanitize [0]{\begingroup\catcode`\&12\catcode`\#12\relax}%
\@ifxundefined \pdfoutput {\@firstoftwo}{%
 \@ifnum{\z@=\pdfoutput}{\@firstoftwo}{\@secondoftwo}%
}{%
 \providecommand\@@startlink[1]{\leavevmode\special{html:<a href="#1">}}%
 \providecommand\@@endlink[0]{\special{html:</a>}}%
}{%
 \providecommand\@@startlink[1]{%
  \leavevmode
  \pdfstartlink
   attr{/Border[0 0 1 ]/H/I/C[0 1 1]}%
   user{/Subtype/Link/A<</Type/Action/S/URI/URI(#1)>>}%
  \relax
 }%
 \providecommand\@@endlink[0]{\pdfendlink}%
}%
\providecommand \url  [0]{\begingroup\@sanitize \@url }%
\providecommand \@url [1]{\endgroup\@href {#1}{\urlprefix}}%
\providecommand \urlprefix [0]{URL }%
\providecommand \Eprint[0]{\href }%
\@ifxundefined \urlstyle {%
  \providecommand \doi [1]{doi:\discretionary{}{}{}#1}%
}{%
  \providecommand \doi [0]{doi:\discretionary{}{}{}\begingroup
  \urlstyle{rm}\Url }%
}%
\providecommand \doibase [0]{http://dx.doi.org/}%
\providecommand \Doi[1]{\href{\doibase#1}}%
\providecommand \selectlanguage [0]{\@gobble}%
\providecommand \bibinfo [0]{\@secondoftwo}%
\providecommand \bibfield [0]{\@secondoftwo}%
\providecommand \translation [1]{[#1]}%
\providecommand \BibitemOpen[0]{}%
\providecommand \bibitemStop [0]{}%
\providecommand \bibitemNoStop [0]{.\EOS\space}%
\providecommand \EOS [0]{\spacefactor3000\relax}%
\providecommand \BibitemShut [1]{\csname bibitem#1\endcsname}%
\bibitem{Davidson:book:TurbulenceRotating}%
  \BibitemOpen
  \bibfield{author}{%
  \bibinfo {author} {\bibfnamefont{P.~A.}\ \bibnamefont{Davidson}},\ }%
  \emph{\bibinfo {title} {{Turbulence in Rotating, Stratified and Electrically
  Conducting Fluids}}}\ (\bibinfo {publisher} {Cambdrige University Press},\
  \bibinfo {address} {Cambdrige(UK)},\ \bibinfo {year}
  {2013})\BibitemShut{NoStop}%
\bibitem{Kolmogorov:DANS1941Structure}%
  \BibitemOpen
  \bibfield{author}{%
  \bibinfo {author} {\bibfnamefont{A.~N.}\ \bibnamefont{Kolmogorov}},\ }%
  \bibfield{title}{%
  \enquote{\bibinfo {title} {{The local structure of turbulence in
  incompressible viscous fluid for very large Reynolds numbers}},}\ }%
  \bibfield{journal}{%
  \bibinfo {journal} {Dokl Acad Nauk SSSR}\ }%
  \textbf{\bibinfo {volume} {30}},\ \bibinfo {pages} {301--305} (\bibinfo
  {year} {1941})\BibitemShut{NoStop}%
\bibitem{Kolmogorov:DANS1941Dissipation}%
  \BibitemOpen
  \bibfield{author}{%
  \bibinfo {author} {\bibfnamefont{A.~N.}\ \bibnamefont{Kolmogorov}},\ }%
  \bibfield{title}{%
  \enquote{\bibinfo {title} {{Dissipation of Energy in Locally Isotropic
  Turbulence}},}\ }%
  \bibfield{journal}{%
  \bibinfo {journal} {Dokl Acad Nauk SSSR}\ }%
  \textbf{\bibinfo {volume} {32}},\ \bibinfo {pages} {16--18} (\bibinfo {year}
  {1941})\BibitemShut{NoStop}%
\bibitem{Elsinga:PF2016}%
  \BibitemOpen
  \bibfield{author}{%
  \bibinfo {author} {\bibfnamefont{G.~E.}\ \bibnamefont{Elsinga}}\ and\
  \bibinfo {author} {\bibfnamefont{Ivan}\ \bibnamefont{Marusic}},\ }%
  \bibfield{title}{%
  \enquote{\bibinfo {title} {{The anisotropic structure of turbulence and its
  energy spectrum}},}\ }%
  \bibfield{journal}{%
  \bibinfo {journal} {Phys. Fluids}\ }%
  \textbf{\bibinfo {volume} {28}},\ \bibinfo {pages} {011701} (\bibinfo {year}
  {2016})\BibitemShut{NoStop}%
\bibitem{Valente:PF2016}%
  \BibitemOpen
  \bibfield{author}{%
  \bibinfo {author} {\bibfnamefont{P.~C.}\ \bibnamefont{Valente}}, \bibinfo
  {author} {\bibfnamefont{C.~B.}\ \bibnamefont{da~Silva}},\ and\ \bibinfo
  {author} {\bibfnamefont{F.~T.}\ \bibnamefont{Pinho}},\ }%
  \bibfield{title}{%
  \enquote{\bibinfo {title} {Energy spectra in elasto-inertial turbulence},}\
  }%
  \bibfield{journal}{%
  \bibinfo {journal} {Phys. Fluids}\ }%
  \textbf{\bibinfo {volume} {28}},\ \bibinfo {pages} {075108} (\bibinfo {year}
  {2016})\BibitemShut{NoStop}%
\bibitem{Pawar:PF2016}%
  \BibitemOpen
  \bibfield{author}{%
  \bibinfo {author} {\bibfnamefont{S.~S.}\ \bibnamefont{Pawar}}\ and\ \bibinfo
  {author} {\bibfnamefont{J.~H.}\ \bibnamefont{Arakeri}},\ }%
  \bibfield{title}{%
  \enquote{\bibinfo {title} {Kinetic energy and scalar spectra in high rayleigh
  number axially homogeneous buoyancy driven turbulence},}\ }%
  \bibfield{journal}{%
  \bibinfo {journal} {Phys. Fluids}\ }%
  \textbf{\bibinfo {volume} {28}},\ \bibinfo {pages} {065103} (\bibinfo {year}
  {2016})\BibitemShut{NoStop}%
\bibitem{Verma:book:BDF}%
  \BibitemOpen
  \bibfield{author}{%
  \bibinfo {author} {\bibfnamefont{M.~K.}\ \bibnamefont{Verma}},\ }%
  \emph{\bibinfo {title} {{Physics of Buoyant Flows, From Instabilities to
  Turbulence}}}\ (\bibinfo {publisher} {World Scientific, Singapore},\ \bibinfo
  {address} {Singapore},\ \bibinfo {year} {2018})\BibitemShut{NoStop}%
\bibitem{Buchhave:PF2017}%
  \BibitemOpen
  \bibfield{author}{%
  \bibinfo {author} {\bibfnamefont{P.}~\bibnamefont{Buchhave}}\ and\ \bibinfo
  {author} {\bibfnamefont{C.~M.}\ \bibnamefont{Velte}},\ }%
  \bibfield{title}{%
  \enquote{\bibinfo {title} {Measurement of turbulent spatial structure and
  kinetic energy spectrum by exact temporal-to-spatial mapping},}\ }%
  \bibfield{journal}{%
  \bibinfo {journal} {Phys. Fluids}\ }%
  \textbf{\bibinfo {volume} {29}},\ \bibinfo {pages} {085109} (\bibinfo {year}
  {2017})\BibitemShut{NoStop}%
\bibitem{Kraichnan:PF1967_2D}%
  \BibitemOpen
  \bibfield{author}{%
  \bibinfo {author} {\bibfnamefont{R.~H.}\ \bibnamefont{Kraichnan}},\ }%
  \bibfield{title}{%
  \enquote{\bibinfo {title} {{Inertial ranges in two-dimensional
  turbulence}},}\ }%
  \bibfield{journal}{%
  \bibinfo {journal} {Phys. Fluids}\ }%
  \textbf{\bibinfo {volume} {10}},\ \bibinfo {pages} {1417} (\bibinfo {year}
  {1967})\BibitemShut{NoStop}%
\bibitem{Paret:PRL1997}%
  \BibitemOpen
  \bibfield{author}{%
  \bibinfo {author} {\bibfnamefont{J.}~\bibnamefont{Paret}}\ and\ \bibinfo
  {author} {\bibfnamefont{P.}~\bibnamefont{Tabeling}},\ }%
  \bibfield{title}{%
  \enquote{\bibinfo {title} {{Experimental observation of the two-dimensional
  inverse energy cascade}},}\ }%
  \bibfield{journal}{%
  \bibinfo {journal} {Phys. Rev. Lett.}\ }%
  \textbf{\bibinfo {volume} {79}},\ \bibinfo {pages} {4162} (\bibinfo {year}
  {1997})\BibitemShut{NoStop}%
\bibitem{Chen:PRL2006}%
  \BibitemOpen
  \bibfield{author}{%
  \bibinfo {author} {\bibfnamefont{S.}~\bibnamefont{Chen}}, \bibinfo {author}
  {\bibfnamefont{R.~E}\ \bibnamefont{Ecke}}, \bibinfo {author}
  {\bibfnamefont{G.~L}\ \bibnamefont{Eyink}}, \bibinfo {author}
  {\bibfnamefont{M.}~\bibnamefont{Rivera}}, \bibinfo {author}
  {\bibfnamefont{M.}~\bibnamefont{Wan}},\ and\ \bibinfo {author}
  {\bibfnamefont{Z.}~\bibnamefont{Xiao}},\ }%
  \bibfield{title}{%
  \enquote{\bibinfo {title} {{Physical mechanism of the two-dimensional inverse
  energy cascade}},}\ }%
  \bibfield{journal}{%
  \bibinfo {journal} {Phys. Rev. Lett.}\ }%
  \textbf{\bibinfo {volume} {96}},\ \bibinfo {pages} {084502} (\bibinfo {year}
  {2006})\BibitemShut{NoStop}%
\bibitem{Vallgren:JFM2011}%
  \BibitemOpen
  \bibfield{author}{%
  \bibinfo {author} {\bibfnamefont{A.}~\bibnamefont{Vallgren}}\ and\ \bibinfo
  {author} {\bibfnamefont{E.}~\bibnamefont{Lindborg}},\ }%
  \bibfield{title}{%
  \enquote{\bibinfo {title} {{The enstrophy cascade in forced two-dimensional
  turbulence}},}\ }%
  \bibfield{journal}{%
  \bibinfo {journal} {J. Fluid Mech.}\ }%
  \textbf{\bibinfo {volume} {671}},\ \bibinfo {pages} {168--183} (\bibinfo
  {year} {2011})\BibitemShut{NoStop}%
\bibitem{Boffetta:ARFM2012}%
  \BibitemOpen
  \bibfield{author}{%
  \bibinfo {author} {\bibfnamefont{G.}~\bibnamefont{Boffetta}}\ and\ \bibinfo
  {author} {\bibfnamefont{R.~E}\ \bibnamefont{Ecke}},\ }%
  \bibfield{title}{%
  \enquote{\bibinfo {title} {{Two-Dimensional Turbulence}},}\ }%
  \bibfield{journal}{%
  \bibinfo {journal} {Annu. Rev. Fluid Mech.}\ }%
  \textbf{\bibinfo {volume} {44}},\ \bibinfo {pages} {427--451} (\bibinfo
  {year} {2012})\BibitemShut{NoStop}%
\bibitem{Ibbetson:JFM1975}%
  \BibitemOpen
  \bibfield{author}{%
  \bibinfo {author} {\bibfnamefont{A.}~\bibnamefont{Ibbetson}}\ and\ \bibinfo
  {author} {\bibfnamefont{D.~J.}\ \bibnamefont{Tritton}},\ }%
  \bibfield{title}{%
  \enquote{\bibinfo {title} {{Experiments on turbulence in a rotating
  fluid}},}\ }%
  \bibfield{journal}{%
  \bibinfo {journal} {J. Fluid Mech.}\ }%
  \textbf{\bibinfo {volume} {68}},\ \bibinfo {pages} {639--672} (\bibinfo
  {year} {1975})\BibitemShut{NoStop}%
\bibitem{Hopfinger:JFM1982}%
  \BibitemOpen
  \bibfield{author}{%
  \bibinfo {author} {\bibfnamefont{E.~J.}\ \bibnamefont{{Hopfinger}}}, \bibinfo
  {author} {\bibfnamefont{F.~K.}\ \bibnamefont{{Browand}}},\ and\ \bibinfo
  {author} {\bibfnamefont{Y.}~\bibnamefont{{Gagne}}},\ }%
  \bibfield{title}{%
  \enquote{\bibinfo {title} {{Turbulence and waves in a rotating tank}},}\ }%
  \bibfield{journal}{%
  \bibinfo {journal} {J. Fluid Mech.}\ }%
  \textbf{\bibinfo {volume} {125}},\ \bibinfo {pages} {505--534} (\bibinfo
  {year} {1982})\BibitemShut{NoStop}%
\bibitem{Bardina:JFM1985}%
  \BibitemOpen
  \bibfield{author}{%
  \bibinfo {author} {\bibfnamefont{J.}~\bibnamefont{Bardina}}, \bibinfo
  {author} {\bibfnamefont{J.~H.}\ \bibnamefont{Ferziger}},\ and\ \bibinfo
  {author} {\bibfnamefont{R.~S.}\ \bibnamefont{Rogallo}},\ }%
  \bibfield{title}{%
  \enquote{\bibinfo {title} {{Effect of rotation on isotropic turbulence -
  Computation and modelling}},}\ }%
  \bibfield{journal}{%
  \bibinfo {journal} {J. Fluid Mech.}\ }%
  \textbf{\bibinfo {volume} {154}},\ \bibinfo {pages} {321--336} (\bibinfo
  {year} {1985})\BibitemShut{NoStop}%
\bibitem{Jacquin:JFM1990}%
  \BibitemOpen
  \bibfield{author}{%
  \bibinfo {author} {\bibfnamefont{L.}~\bibnamefont{Jacquin}}, \bibinfo
  {author} {\bibfnamefont{O.}~\bibnamefont{Leuchter}}, \bibinfo {author}
  {\bibfnamefont{C.}~\bibnamefont{Cambon}},\ and\ \bibinfo {author}
  {\bibfnamefont{J.}~\bibnamefont{Mathieu}},\ }%
  \bibfield{title}{%
  \enquote{\bibinfo {title} {{Homogeneous turbulence in the presence of
  rotation}},}\ }%
  \bibfield{journal}{%
  \bibinfo {journal} {J. Fluid Mech.}\ }%
  \textbf{\bibinfo {volume} {220}},\ \bibinfo {pages} {1--52} (\bibinfo {year}
  {1990})\BibitemShut{NoStop}%
\bibitem{Zeman:PF1994}%
  \BibitemOpen
  \bibfield{author}{%
  \bibinfo {author} {\bibfnamefont{O.}~\bibnamefont{Zeman}},\ }%
  \bibfield{title}{%
  \enquote{\bibinfo {title} {{A note on the spectra and decay of rotating
  homogeneous turbulence}},}\ }%
  \bibfield{journal}{%
  \bibinfo {journal} {Phys. Fluids}\ }%
  \textbf{\bibinfo {volume} {6}},\ \bibinfo {pages} {3221--3224} (\bibinfo
  {year} {1994})\BibitemShut{NoStop}%
\bibitem{Zhou:PF1995}%
  \BibitemOpen
  \bibfield{author}{%
  \bibinfo {author} {\bibfnamefont{Y.}~\bibnamefont{Zhou}},\ }%
  \bibfield{title}{%
  \enquote{\bibinfo {title} {{A phenomenological treatment of rotating
  turbulence}},}\ }%
  \bibfield{journal}{%
  \bibinfo {journal} {Phys. Fluids}\ }%
  \textbf{\bibinfo {volume} {7}},\ \bibinfo {pages} {2092} (\bibinfo {year}
  {1995})\BibitemShut{NoStop}%
\bibitem{Canuto:PRL1997}%
  \BibitemOpen
  \bibfield{author}{%
  \bibinfo {author} {\bibfnamefont{V.~M.}\ \bibnamefont{Canuto}}\ and\ \bibinfo
  {author} {\bibfnamefont{M.~S.}\ \bibnamefont{Dubovikov}},\ }%
  \bibfield{title}{%
  \enquote{\bibinfo {title} {{Physical regimes and dimensional structure of
  rotating turbulence}},}\ }%
  \bibfield{journal}{%
  \bibinfo {journal} {Phys. Rev. Lett.}\ }%
  \textbf{\bibinfo {volume} {78}},\ \bibinfo {pages} {666--669} (\bibinfo
  {year} {1997})\BibitemShut{NoStop}%
\bibitem{Canuto:PF1997_rotation}%
  \BibitemOpen
  \bibfield{author}{%
  \bibinfo {author} {\bibfnamefont{V.~M.}\ \bibnamefont{Canuto}}\ and\ \bibinfo
  {author} {\bibfnamefont{M.~S.}\ \bibnamefont{Dubovikov}},\ }%
  \bibfield{title}{%
  \enquote{\bibinfo {title} {{A dynamical model for turbulence. V. The effect
  of rotation}},}\ }%
  \bibfield{journal}{%
  \bibinfo {journal} {Phys. Fluids}\ }%
  \textbf{\bibinfo {volume} {9}},\ \bibinfo {pages} {2132--2140} (\bibinfo
  {year} {1997})\BibitemShut{NoStop}%
\bibitem{Yeung:PF1998}%
  \BibitemOpen
  \bibfield{author}{%
  \bibinfo {author} {\bibfnamefont{P.~K.}\ \bibnamefont{Yeung}}\ and\ \bibinfo
  {author} {\bibfnamefont{Y.}~\bibnamefont{Zhou}},\ }%
  \bibfield{title}{%
  \enquote{\bibinfo {title} {{Numerical study of rotating turbulence with
  external forcing}},}\ }%
  \bibfield{journal}{%
  \bibinfo {journal} {Phys. Fluids}\ }%
  \textbf{\bibinfo {volume} {10}},\ \bibinfo {pages} {2895--2909} (\bibinfo
  {year} {1998})\BibitemShut{NoStop}%
\bibitem{Smith:PF1999}%
  \BibitemOpen
  \bibfield{author}{%
  \bibinfo {author} {\bibfnamefont{L.~M.}\ \bibnamefont{Smith}}\ and\ \bibinfo
  {author} {\bibfnamefont{F.}~\bibnamefont{Waleffe}},\ }%
  \bibfield{title}{%
  \enquote{\bibinfo {title} {{Transfer of energy to two-dimensional large
  scales in forced, rotating three-dimensional turbulence}},}\ }%
  \bibfield{journal}{%
  \bibinfo {journal} {Phys. Fluids}\ }%
  \textbf{\bibinfo {volume} {11}},\ \bibinfo {pages} {1608} (\bibinfo {year}
  {1999})\BibitemShut{NoStop}%
\bibitem{Baroud:PRL2002}%
  \BibitemOpen
  \bibfield{author}{%
  \bibinfo {author} {\bibfnamefont{C.~N.}\ \bibnamefont{{Baroud}}}, \bibinfo
  {author} {\bibfnamefont{B.~B.}\ \bibnamefont{{Plapp}}}, \bibinfo {author}
  {\bibfnamefont{Z.-S.}\ \bibnamefont{{She}}},\ and\ \bibinfo {author}
  {\bibfnamefont{H.~L.}\ \bibnamefont{{Swinney}}},\ }%
  \bibfield{title}{%
  \enquote{\bibinfo {title} {Anomalous self-similarity in a turbulent rapidly
  rotating fluid},}\ }%
  \bibfield{journal}{%
  \Doi{10.1103/PhysRevLett.88.114501}{\bibinfo {journal} {Phys. Rev. Lett.}}\
  }%
  \textbf{\bibinfo {volume} {88}},\ \bibinfo {eid} {114501} (\bibinfo {year}
  {2002})\BibitemShut{NoStop}%
\bibitem{Galtier:PRE2003}%
  \BibitemOpen
  \bibfield{author}{%
  \bibinfo {author} {\bibfnamefont{S.}~\bibnamefont{Galtier}},\ }%
  \bibfield{title}{%
  \enquote{\bibinfo {title} {{Weak inertial-wave turbulence theory}},}\ }%
  \bibfield{journal}{%
  \bibinfo {journal} {Phys. Rev. E}\ }%
  \textbf{\bibinfo {volume} {68}},\ \bibinfo {pages} {015301--4} (\bibinfo
  {year} {2003})\BibitemShut{NoStop}%
\bibitem{Hattori:PRE2004}%
  \BibitemOpen
  \bibfield{author}{%
  \bibinfo {author} {\bibfnamefont{Y.}~\bibnamefont{Hattori}}, \bibinfo
  {author} {\bibfnamefont{R.}~\bibnamefont{Rubinstein}},\ and\ \bibinfo
  {author} {\bibfnamefont{A.}~\bibnamefont{Ishizawa}},\ }%
  \bibfield{title}{%
  \enquote{\bibinfo {title} {Shell model for rotating turbulence},}\ }%
  \bibfield{journal}{%
  \bibinfo {journal} {Phys. Rev. E}\ }%
  \textbf{\bibinfo {volume} {70}},\ \bibinfo {pages} {046311} (\bibinfo {year}
  {2004})\BibitemShut{NoStop}%
\bibitem{Yang:PF2004}%
  \BibitemOpen
  \bibfield{author}{%
  \bibinfo {author} {\bibfnamefont{X.}~\bibnamefont{Yang}}\ and\ \bibinfo
  {author} {\bibfnamefont{J.~A.}\ \bibnamefont{Domaradzki}},\ }%
  \bibfield{title}{%
  \enquote{\bibinfo {title} {{Large eddy simulations of decaying rotating
  turbulence}},}\ }%
  \bibfield{journal}{%
  \bibinfo {journal} {Phys. Fluids}\ }%
  \textbf{\bibinfo {volume} {16}},\ \bibinfo {pages} {4088--4104} (\bibinfo
  {year} {2004})\BibitemShut{NoStop}%
\bibitem{Felsot:PRE2005}%
  \BibitemOpen
  \bibfield{author}{%
  \bibinfo {author} {\bibfnamefont{J.~E.}\ \bibnamefont{Ruppert-Felsot}},
  \bibinfo {author} {\bibfnamefont{O.}~\bibnamefont{Praud}}, \bibinfo {author}
  {\bibfnamefont{E.}~\bibnamefont{Sharon}},\ and\ \bibinfo {author}
  {\bibfnamefont{H.~L.}\ \bibnamefont{Swinney}},\ }%
  \bibfield{title}{%
  \enquote{\bibinfo {title} {Extraction of coherent structures in a rotating
  turbulent flow experiment},}\ }%
  \bibfield{journal}{%
  \bibinfo {journal} {Phys. Rev. E}\ }%
  \textbf{\bibinfo {volume} {72}},\ \bibinfo {pages} {016311} (\bibinfo {year}
  {2005})\BibitemShut{NoStop}%
\bibitem{Smith:JFM2005}%
  \BibitemOpen
  \bibfield{author}{%
  \bibinfo {author} {\bibfnamefont{L.~M.}\ \bibnamefont{Smith}}\ and\ \bibinfo
  {author} {\bibfnamefont{Y.}~\bibnamefont{Lee}},\ }%
  \bibfield{title}{%
  \enquote{\bibinfo {title} {{On near resonances and symmetry breaking in
  forced rotating flows at moderate Rossby number}},}\ }%
  \bibfield{journal}{%
  \bibinfo {journal} {J. Fluid Mech.}\ }%
  \textbf{\bibinfo {volume} {535}},\ \bibinfo {pages} {111--142} (\bibinfo
  {year} {2005})\BibitemShut{NoStop}%
\bibitem{Morize:PF2005}%
  \BibitemOpen
  \bibfield{author}{%
  \bibinfo {author} {\bibfnamefont{C.}~\bibnamefont{Morize}}, \bibinfo {author}
  {\bibfnamefont{F.}~\bibnamefont{Moisy}},\ and\ \bibinfo {author}
  {\bibfnamefont{M.}~\bibnamefont{Rabaud}},\ }%
  \bibfield{title}{%
  \enquote{\bibinfo {title} {{Decaying grid-generated turbulence in a rotating
  tank}},}\ }%
  \bibfield{journal}{%
  \bibinfo {journal} {Phys. Fluids}\ }%
  \textbf{\bibinfo {volume} {17}},\ \bibinfo {pages} {095105} (\bibinfo {year}
  {2005})\BibitemShut{NoStop}%
\bibitem{Muller:EPL2007}%
  \BibitemOpen
  \bibfield{author}{%
  \bibinfo {author} {\bibfnamefont{W.-C.}\ \bibnamefont{M{\"u}ller}}\ and\
  \bibinfo {author} {\bibfnamefont{M}~\bibnamefont{Thiele}},\ }%
  \bibfield{title}{%
  \enquote{\bibinfo {title} {{Scaling and energy transfer in rotating
  turbulence}},}\ }%
  \bibfield{journal}{%
  \bibinfo {journal} {EPL}\ }%
  \textbf{\bibinfo {volume} {77}},\ \bibinfo {pages} {34003} (\bibinfo {year}
  {2007})\BibitemShut{NoStop}%
\bibitem{Chakraborty:PRE2007}%
  \BibitemOpen
  \bibfield{author}{%
  \bibinfo {author} {\bibfnamefont{S.}~\bibnamefont{Chakraborty}}\ and\
  \bibinfo {author} {\bibfnamefont{J.~K.}\ \bibnamefont{Bhattacharjee}},\ }%
  \bibfield{title}{%
  \enquote{\bibinfo {title} {{Third-order structure function for rotating
  three-dimensional homogeneous turbulent flow}},}\ }%
  \bibfield{journal}{%
  \bibinfo {journal} {Phys. Rev. E}\ }%
  \textbf{\bibinfo {volume} {76}},\ \bibinfo {pages} {036304} (\bibinfo {year}
  {2007})\BibitemShut{NoStop}%
\bibitem{Chakraborty:EPL2007}%
  \BibitemOpen
  \bibfield{author}{%
  \bibinfo {author} {\bibfnamefont{S.}~\bibnamefont{Chakraborty}},\ }%
  \bibfield{title}{%
  \enquote{\bibinfo {title} {{Signatures of two-dimensionalisation of 3D
  turbulence in the presence of rotation}},}\ }%
  \bibfield{journal}{%
  \bibinfo {journal} {EPL}\ }%
  \textbf{\bibinfo {volume} {79}},\ \bibinfo {pages} {14002} (\bibinfo {year}
  {2007})\BibitemShut{NoStop}%
\bibitem{Thiele:JFM2009}%
  \BibitemOpen
  \bibfield{author}{%
  \bibinfo {author} {\bibfnamefont{M.}~\bibnamefont{Thiele}}\ and\ \bibinfo
  {author} {\bibfnamefont{W.-C.}\ \bibnamefont{M{\"u}ller}},\ }%
  \bibfield{title}{%
  \enquote{\bibinfo {title} {{Structure and decay of rotating homogeneous
  turbulence}},}\ }%
  \bibfield{journal}{%
  \bibinfo {journal} {J. Fluid Mech.}\ }%
  \textbf{\bibinfo {volume} {637}},\ \bibinfo {pages} {425--442} (\bibinfo
  {year} {2009})\BibitemShut{NoStop}%
\bibitem{Mininni:PF2009}%
  \BibitemOpen
  \bibfield{author}{%
  \bibinfo {author} {\bibfnamefont{P.~D.}\ \bibnamefont{Mininni}}, \bibinfo
  {author} {\bibfnamefont{A.}~\bibnamefont{Alexakis}},\ and\ \bibinfo {author}
  {\bibfnamefont{A.}~\bibnamefont{Pouquet}},\ }%
  \bibfield{title}{%
  \enquote{\bibinfo {title} {{Scale interactions and scaling laws in rotating
  flows at moderate Rossby numbers and large Reynolds numbers}},}\ }%
  \bibfield{journal}{%
  \bibinfo {journal} {Phys. Fluids}\ }%
  \textbf{\bibinfo {volume} {21}},\ \bibinfo {pages} {015108} (\bibinfo {year}
  {2009})\BibitemShut{NoStop}%
\bibitem{Chakraborty:EPJB2010}%
  \BibitemOpen
  \bibfield{author}{%
  \bibinfo {author} {\bibfnamefont{S.}~\bibnamefont{Chakraborty}}, \bibinfo
  {author} {\bibfnamefont{M.~H.}\ \bibnamefont{Jensen}},\ and\ \bibinfo
  {author} {\bibfnamefont{A.}~\bibnamefont{Sarkar}},\ }%
  \bibfield{title}{%
  \enquote{\bibinfo {title} {{On two-dimensionalization of three-dimensional
  turbulence in shell models}},}\ }%
  \bibfield{journal}{%
  \bibinfo {journal} {Eur. Phys. J. B}\ }%
  \textbf{\bibinfo {volume} {73}},\ \bibinfo {pages} {447--453} (\bibinfo
  {year} {2010})\BibitemShut{NoStop}%
\bibitem{Sen:PRE2012}%
  \BibitemOpen
  \bibfield{author}{%
  \bibinfo {author} {\bibfnamefont{A.}~\bibnamefont{Sen}}, \bibinfo {author}
  {\bibfnamefont{P.~D.}\ \bibnamefont{Mininni}}, \bibinfo {author}
  {\bibfnamefont{D.}~\bibnamefont{Rosenberg}},\ and\ \bibinfo {author}
  {\bibfnamefont{A.}~\bibnamefont{Pouquet}},\ }%
  \bibfield{title}{%
  \enquote{\bibinfo {title} {{Anisotropy and nonuniversality in scaling laws of
  the large-scale energy spectrum in rotating turbulence}},}\ }%
  \bibfield{journal}{%
  \bibinfo {journal} {Phys. Rev. E}\ }%
  \textbf{\bibinfo {volume} {86}},\ \bibinfo {pages} {036319} (\bibinfo {year}
  {2012})\BibitemShut{NoStop}%
\bibitem{Biferale:PRX2016}%
  \BibitemOpen
  \bibfield{author}{%
  \bibinfo {author} {\bibfnamefont{L.}~\bibnamefont{Biferale}}, \bibinfo
  {author} {\bibfnamefont{F.}~\bibnamefont{Bonaccorso}}, \bibinfo {author}
  {\bibfnamefont{I.~M.}\ \bibnamefont{Mazzitelli}}, \bibinfo {author}
  {\bibfnamefont{M.~A.~T.}\ \bibnamefont{van Hinsberg}}, \bibinfo {author}
  {\bibfnamefont{A.~S.}\ \bibnamefont{Lanotte}}, \bibinfo {author}
  {\bibfnamefont{S.}~\bibnamefont{Musacchio}}, \bibinfo {author}
  {\bibfnamefont{P}~\bibnamefont{Perlekar}},\ and\ \bibinfo {author}
  {\bibfnamefont{F.}~\bibnamefont{Toschi}},\ }%
  \bibfield{title}{%
  \enquote{\bibinfo {title} {{Coherent Structures and Extreme Events in
  Rotating Multiphase Turbulent Flows}},}\ }%
  \bibfield{journal}{%
  \bibinfo {journal} {Phys. Rev. X}\ }%
  \textbf{\bibinfo {volume} {6}},\ \bibinfo {pages} {041036} (\bibinfo {year}
  {2016})\BibitemShut{NoStop}%
\bibitem{Baqui:PF2016}%
  \BibitemOpen
  \bibfield{author}{%
  \bibinfo {author} {\bibfnamefont{Y.~B.}\ \bibnamefont{Baqui}}, \bibinfo
  {author} {\bibfnamefont{P.~A.}\ \bibnamefont{Davidson}},\ and\ \bibinfo
  {author} {\bibfnamefont{A.}~\bibnamefont{Ranjan}},\ }%
  \bibfield{title}{%
  \enquote{\bibinfo {title} {{Are there two regimes in strongly rotating
  turbulence?}}.}\ }%
  \bibfield{journal}{%
  \bibinfo {journal} {Phys. Fluids}\ }%
  \textbf{\bibinfo {volume} {28}},\ \bibinfo {pages} {045103} (\bibinfo {year}
  {2016})\BibitemShut{NoStop}%
\bibitem{Hossain:PF1994}%
  \BibitemOpen
  \bibfield{author}{%
  \bibinfo {author} {\bibfnamefont{M.}~\bibnamefont{Hossain}},\ }%
  \bibfield{title}{%
  \enquote{\bibinfo {title} {{Reduction in the dimensionality of turbulence due
  to a strong rotation}},}\ }%
  \bibfield{journal}{%
  \bibinfo {journal} {Phys. Fluids}\ }%
  \textbf{\bibinfo {volume} {6}},\ \bibinfo {pages} {1077--1080} (\bibinfo
  {year} {1994})\BibitemShut{NoStop}%
\bibitem{Smith:PRL1996_cross}%
  \BibitemOpen
  \bibfield{author}{%
  \bibinfo {author} {\bibfnamefont{L.~M.}\ \bibnamefont{Smith}}, \bibinfo
  {author} {\bibfnamefont{J.~R.}\ \bibnamefont{Chasnov}},\ and\ \bibinfo
  {author} {\bibfnamefont{F.}~\bibnamefont{Waleffe}},\ }%
  \bibfield{title}{%
  \enquote{\bibinfo {title} {Crossover from two- to three-dimensional
  turbulence},}\ }%
  \bibfield{journal}{%
  \bibinfo {journal} {Phys. Rev. Lett.}\ }%
  \textbf{\bibinfo {volume} {77}},\ \bibinfo {pages} {2467--2470} (\bibinfo
  {year} {1996})\BibitemShut{NoStop}%
\bibitem{Poquet:IOP2013}%
  \BibitemOpen
  \bibfield{author}{%
  \bibinfo {author} {\bibfnamefont{A.}~\bibnamefont{Pouquet}}, \bibinfo
  {author} {\bibfnamefont{A.}~\bibnamefont{Sen}}, \bibinfo {author}
  {\bibfnamefont{D.}~\bibnamefont{Rosenberg}}, \bibinfo {author}
  {\bibfnamefont{P.~D.}\ \bibnamefont{Mininni}},\ and\ \bibinfo {author}
  {\bibfnamefont{J}~\bibnamefont{Baerenzung}},\ }%
  \bibfield{title}{%
  \enquote{\bibinfo {title} {Inverse cascades in turbulence and the case of
  rotating flows},}\ }%
  \bibfield{journal}{%
  \bibinfo {journal} {Phys. Scr.}\ }%
  \textbf{\bibinfo {volume} {T155}},\ \bibinfo {pages} {014032} (\bibinfo
  {year} {2013})\BibitemShut{NoStop}%
\bibitem{Yarom:PF2013}%
  \BibitemOpen
  \bibfield{author}{%
  \bibinfo {author} {\bibfnamefont{E.}~\bibnamefont{Yarom}}, \bibinfo {author}
  {\bibfnamefont{Y.}~\bibnamefont{Vardi}},\ and\ \bibinfo {author}
  {\bibfnamefont{E.}~\bibnamefont{Sharon}},\ }%
  \bibfield{title}{%
  \enquote{\bibinfo {title} {Experimental quantification of inverse energy
  cascade in deep rotating turbulence},}\ }%
  \bibfield{journal}{%
  \bibinfo {journal} {Phys. Fluids}\ }%
  \textbf{\bibinfo {volume} {25}},\ \bibinfo {pages} {085105} (\bibinfo {year}
  {2013})\BibitemShut{NoStop}%
\bibitem{Campagne:PF2014}%
  \BibitemOpen
  \bibfield{author}{%
  \bibinfo {author} {\bibfnamefont{A.}~\bibnamefont{Campagne}}, \bibinfo
  {author} {\bibfnamefont{B.}~\bibnamefont{Gallet}}, \bibinfo {author}
  {\bibfnamefont{F.}~\bibnamefont{Moisy}},\ and\ \bibinfo {author}
  {\bibfnamefont{P.-P.}\ \bibnamefont{Cortet}},\ }%
  \bibfield{title}{%
  \enquote{\bibinfo {title} {{Direct and inverse energy cascades in a forced
  rotating turbulence experiment}},}\ }%
  \bibfield{journal}{%
  \bibinfo {journal} {Phys. Fluids}\ }%
  \textbf{\bibinfo {volume} {26}},\ \bibinfo {pages} {125112} (\bibinfo {year}
  {2014})\BibitemShut{NoStop}%
\bibitem{Bourouiba:JFM2011}%
  \BibitemOpen
  \bibfield{author}{%
  \bibinfo {author} {\bibfnamefont{L.}~\bibnamefont{Bourouiba}}, \bibinfo
  {author} {\bibfnamefont{D.~N.}\ \bibnamefont{Straub}},\ and\ \bibinfo
  {author} {\bibfnamefont{M.~L.}\ \bibnamefont{Waite}},\ }%
  \bibfield{title}{%
  \enquote{\bibinfo {title} {{Non-local energy transfers in rotating turbulence
  at intermediate Rossby number}},}\ }%
  \bibfield{journal}{%
  \bibinfo {journal} {J. Fluid Mech.}\ }%
  \textbf{\bibinfo {volume} {690}},\ \bibinfo {pages} {129--147} (\bibinfo
  {year} {2011})\BibitemShut{NoStop}%
\bibitem{Kraichnan:JFM1959}%
  \BibitemOpen
  \bibfield{author}{%
  \bibinfo {author} {\bibfnamefont{R.~H.}\ \bibnamefont{Kraichnan}},\ }%
  \bibfield{title}{%
  \enquote{\bibinfo {title} {{The structure of isotropic turbulence at very
  high Reynolds numbers}},}\ }%
  \bibfield{journal}{%
  \bibinfo {journal} {J. Fluid Mech.}\ }%
  \textbf{\bibinfo {volume} {5}},\ \bibinfo {pages} {497--543} (\bibinfo {year}
  {1959})\BibitemShut{NoStop}%
\bibitem{Pao:PF1965}%
  \BibitemOpen
  \bibfield{author}{%
  \bibinfo {author} {\bibfnamefont{Y.-H.}\ \bibnamefont{Pao}},\ }%
  \bibfield{title}{%
  \enquote{\bibinfo {title} {{Structure of Turbulent Velocity and Scalar Fields
  at Large Wavenumbers}},}\ }%
  \bibfield{journal}{%
  \bibinfo {journal} {Phys. Fluids}\ }%
  \textbf{\bibinfo {volume} {8}},\ \bibinfo {pages} {1063} (\bibinfo {year}
  {1965})\BibitemShut{NoStop}%
\bibitem{Pope:book}%
  \BibitemOpen
  \bibfield{author}{%
  \bibinfo {author} {\bibfnamefont{S.~B.}\ \bibnamefont{Pope}},\ }%
  \emph{\bibinfo {title} {{Turbulent Flows}}}\ (\bibinfo {publisher} {Cambridge
  University Press},\ \bibinfo {address} {Cambridge},\ \bibinfo {year}
  {2000})\BibitemShut{NoStop}%
\bibitem{Smith:PF1991}%
  \BibitemOpen
  \bibfield{author}{%
  \bibinfo {author} {\bibfnamefont{L.~M.}\ \bibnamefont{Smith}}\ and\ \bibinfo
  {author} {\bibfnamefont{W.~C.}\ \bibnamefont{Reynolds}},\ }%
  \bibfield{title}{%
  \enquote{\bibinfo {title} {{The dissipation-range spectrum and the
  velocity-derivative skewness in turbulent flows}},}\ }%
  \bibfield{journal}{%
  \bibinfo {journal} {Phys. Fluids A}\ }%
  \textbf{\bibinfo {volume} {3}},\ \bibinfo {pages} {992} (\bibinfo {year}
  {1991})\BibitemShut{NoStop}%
\bibitem{Sreenivasan:JFM1985}%
  \BibitemOpen
  \bibfield{author}{%
  \bibinfo {author} {\bibfnamefont{K.~R.}\ \bibnamefont{Sreenivasan}},\ }%
  \bibfield{title}{%
  \enquote{\bibinfo {title} {{On the fine-scale intermittency of
  turbulence}},}\ }%
  \bibfield{journal}{%
  \bibinfo {journal} {J. Fluid Mech.}\ }%
  \textbf{\bibinfo {volume} {151}},\ \bibinfo {pages} {81--103} (\bibinfo
  {year} {1985})\BibitemShut{NoStop}%
\bibitem{Foias:PF1990}%
  \BibitemOpen
  \bibfield{author}{%
  \bibinfo {author} {\bibfnamefont{C.}~\bibnamefont{{Foias}}}, \bibinfo
  {author} {\bibfnamefont{O.}~\bibnamefont{{Manley}}},\ and\ \bibinfo {author}
  {\bibfnamefont{L.}~\bibnamefont{{Sirovich}}},\ }%
  \bibfield{title}{%
  \enquote{\bibinfo {title} {Empirical and stokes eignfunctions and
  far-dissipative turbulent spectrum},}\ }%
  \bibfield{journal}{%
  \Doi{10.1063/1.857744}{\bibinfo {journal} {Phys. Fluids A}}\ }%
  \textbf{\bibinfo {volume} {2}},\ \bibinfo {pages} {464} (\bibinfo {year}
  {1990})\BibitemShut{NoStop}%
\bibitem{Sanada:PF1992d}%
  \BibitemOpen
  \bibfield{author}{%
  \bibinfo {author} {\bibfnamefont{T.}~\bibnamefont{Sanada}},\ }%
  \bibfield{title}{%
  \enquote{\bibinfo {title} {{Comment on the dssipation-range spectrum in
  turbulent flows}},}\ }%
  \bibfield{journal}{%
  \bibinfo {journal} {Phys. Fluids A}\ }%
  \textbf{\bibinfo {volume} {4}},\ \bibinfo {pages} {1086} (\bibinfo {year}
  {1992})\BibitemShut{NoStop}%
\bibitem{Manley:PF1992}%
  \BibitemOpen
  \bibfield{author}{%
  \bibinfo {author} {\bibfnamefont{O.~P.}\ \bibnamefont{{Manley}}},\ }%
  \bibfield{title}{%
  \enquote{\bibinfo {title} {The dissipation range spectrum},}\ }%
  \bibfield{journal}{%
  \Doi{10.1063/1.858408}{\bibinfo {journal} {Phys. Fluids A}}\ }%
  \textbf{\bibinfo {volume} {4}},\ \bibinfo {pages} {1320} (\bibinfo {year}
  {1992})\BibitemShut{NoStop}%
\bibitem{Martinez:JPP1997}%
  \BibitemOpen
  \bibfield{author}{%
  \bibinfo {author} {\bibfnamefont{D.~O.}\ \bibnamefont{Mart{\'\i}nez}},
  \bibinfo {author} {\bibfnamefont{S.}~\bibnamefont{Chen}}, \bibinfo {author}
  {\bibfnamefont{G.~D.}\ \bibnamefont{Doolen}}, \bibinfo {author}
  {\bibfnamefont{R.~H.}\ \bibnamefont{Kraichnan}}, \bibinfo {author}
  {\bibfnamefont{L.-P.}\ \bibnamefont{Wang}},\ and\ \bibinfo {author}
  {\bibfnamefont{Y.}~\bibnamefont{Zhou}},\ }%
  \bibfield{title}{%
  \enquote{\bibinfo {title} {{Energy spectrum in the dissipation range of fluid
  turbulence}},}\ }%
  \bibfield{journal}{%
  \bibinfo {journal} {J. Plasma Phys.}\ }%
  \textbf{\bibinfo {volume} {57}},\ \bibinfo {pages} {195--201} (\bibinfo
  {year} {1997})\BibitemShut{NoStop}%
\bibitem{Kida:PF1987}%
  \BibitemOpen
  \bibfield{author}{%
  \bibinfo {author} {\bibfnamefont{S.}~\bibnamefont{Kida}}\ and\ \bibinfo
  {author} {\bibfnamefont{Y.}~\bibnamefont{Murakami}},\ }%
  \bibfield{title}{%
  \enquote{\bibinfo {title} {{Kolmogorov similarity in free decaying
  turbulence}},}\ }%
  \bibfield{journal}{%
  \bibinfo {journal} {Phys. Fluids}\ }%
  \textbf{\bibinfo {volume} {30}},\ \bibinfo {pages} {2030} (\bibinfo {year}
  {1987})\BibitemShut{NoStop}%
\bibitem{Domaradzki:PF1992}%
  \BibitemOpen
  \bibfield{author}{%
  \bibinfo {author} {\bibfnamefont{J.~A.}\ \bibnamefont{Domaradzki}},\ }%
  \bibfield{title}{%
  \enquote{\bibinfo {title} {{Nonlocal triad interactions and the dissipation
  range of isotropic turbulence}},}\ }%
  \bibfield{journal}{%
  \bibinfo {journal} {Phys. Fluids A}\ }%
  \textbf{\bibinfo {volume} {4}},\ \bibinfo {pages} {2037} (\bibinfo {year}
  {1992})\BibitemShut{NoStop}%
\bibitem{Chen:PRL1993}%
  \BibitemOpen
  \bibfield{author}{%
  \bibinfo {author} {\bibfnamefont{S.}~\bibnamefont{Chen}}, \bibinfo {author}
  {\bibfnamefont{G.~D.}\ \bibnamefont{Doolen}}, \bibinfo {author}
  {\bibfnamefont{J.~R.}\ \bibnamefont{Herring}}, \bibinfo {author}
  {\bibfnamefont{R.~H.}\ \bibnamefont{Kraichnan}}, \bibinfo {author}
  {\bibfnamefont{S.~A.}\ \bibnamefont{Orszag}},\ and\ \bibinfo {author}
  {\bibfnamefont{Z.-S.}\ \bibnamefont{She}},\ }%
  \bibfield{title}{%
  \enquote{\bibinfo {title} {{Far-dissipation range of turbulence}},}\ }%
  \bibfield{journal}{%
  \bibinfo {journal} {Phys. Rev. Lett.}\ }%
  \textbf{\bibinfo {volume} {70}},\ \bibinfo {pages} {3051} (\bibinfo {year}
  {1993})\BibitemShut{NoStop}%
\bibitem{Saddoughi:JFM1994}%
  \BibitemOpen
  \bibfield{author}{%
  \bibinfo {author} {\bibfnamefont{S.~G.}\ \bibnamefont{Saddoughi}}\ and\
  \bibinfo {author} {\bibfnamefont{S.~V.}\ \bibnamefont{Veeravalli}},\ }%
  \bibfield{title}{%
  \enquote{\bibinfo {title} {{Local isotropy in turbulent boundary layers at
  high Reynolds number}},}\ }%
  \bibfield{journal}{%
  \bibinfo {journal} {J. Fluid Mech.}\ }%
  \textbf{\bibinfo {volume} {268}},\ \bibinfo {pages} {333--372} (\bibinfo
  {year} {1994})\BibitemShut{NoStop}%
\bibitem{Sharma:PF2018}%
  \BibitemOpen
  \bibfield{author}{%
  \bibinfo {author} {\bibfnamefont{M.~K.}\ \bibnamefont{Sharma}}, \bibinfo
  {author} {\bibfnamefont{A.}~\bibnamefont{Kumar}}, \bibinfo {author}
  {\bibfnamefont{M.~K.}\ \bibnamefont{Verma}},\ and\ \bibinfo {author}
  {\bibfnamefont{S.}~\bibnamefont{Chakraborty}},\ }%
  \bibfield{title}{%
  \enquote{\bibinfo {title} {Statistical features of rapidly rotating decaying
  turbulence: Enstrophy and energy spectra and coherent structures},}\ }%
  \bibfield{journal}{%
  \bibinfo {journal} {Phys. Fluids}\ }%
  \textbf{\bibinfo {volume} {30}},\ \bibinfo {pages} {045103} (\bibinfo {year}
  {2018})\BibitemShut{NoStop}%
\bibitem{Verma:Pramana2013tarang}%
  \BibitemOpen
  \bibfield{author}{%
  \bibinfo {author} {\bibfnamefont{M.~K.}\ \bibnamefont{Verma}}, \bibinfo
  {author} {\bibfnamefont{A.~G.}\ \bibnamefont{Chatterjee}}, \bibinfo {author}
  {\bibfnamefont{K.~S.}\ \bibnamefont{Reddy}}, \bibinfo {author}
  {\bibfnamefont{R.~K.}\ \bibnamefont{Yadav}}, \bibinfo {author}
  {\bibfnamefont{S.}~\bibnamefont{Paul}}, \bibinfo {author}
  {\bibfnamefont{M.}~\bibnamefont{Chandra}},\ and\ \bibinfo {author}
  {\bibfnamefont{R.}~\bibnamefont{Samtaney}},\ }%
  \bibfield{title}{%
  \enquote{\bibinfo {title} {{Benchmarking and scaling studies of
  pseudospectral code Tarang for turbulence simulations}},}\ }%
  \bibfield{journal}{%
  \bibinfo {journal} {Pramana-J. Phys.}\ }%
  \textbf{\bibinfo {volume} {81}},\ \bibinfo {pages} {617--629} (\bibinfo
  {year} {2013})\BibitemShut{NoStop}%
\bibitem{Chatterjee:JPDC2018}%
  \BibitemOpen
  \bibfield{author}{%
  \bibinfo {author} {\bibfnamefont{A.~G.}\ \bibnamefont{Chatterjee}}, \bibinfo
  {author} {\bibfnamefont{M.~K.}\ \bibnamefont{Verma}}, \bibinfo {author}
  {\bibfnamefont{A.}~\bibnamefont{Kumar}}, \bibinfo {author}
  {\bibfnamefont{R.}~\bibnamefont{Samtaney}}, \bibinfo {author}
  {\bibfnamefont{B.}~\bibnamefont{Hadri}},\ and\ \bibinfo {author}
  {\bibfnamefont{R.}~\bibnamefont{Khurram}},\ }%
  \bibfield{title}{%
  \enquote{\bibinfo {title} {Scaling of a fast fourier transform and a
  pseudo-spectral fluid solver up to 196608 cores},}\ }%
  \bibfield{journal}{%
  \bibinfo {journal} {J Parallel Distrib Comput}\ }%
  \textbf{\bibinfo {volume} {113}},\ \bibinfo {pages} {77 -- 91} (\bibinfo
  {year} {2018})\BibitemShut{NoStop}%
\bibitem{Dar:PD2001}%
  \BibitemOpen
  \bibfield{author}{%
  \bibinfo {author} {\bibfnamefont{G.}~\bibnamefont{Dar}}, \bibinfo {author}
  {\bibfnamefont{M.~K.}\ \bibnamefont{Verma}},\ and\ \bibinfo {author}
  {\bibfnamefont{V.}~\bibnamefont{Eswaran}},\ }%
  \bibfield{title}{%
  \enquote{\bibinfo {title} {{Energy transfer in two-dimensional
  magnetohydrodynamic turbulence: formalism and numerical results}},}\ }%
  \bibfield{journal}{%
  \bibinfo {journal} {Physica D}\ }%
  \textbf{\bibinfo {volume} {157}},\ \bibinfo {pages} {207--225} (\bibinfo
  {year} {2001})\BibitemShut{NoStop}%
\bibitem{Verma:PR2004}%
  \BibitemOpen
  \bibfield{author}{%
  \bibinfo {author} {\bibfnamefont{M.~K.}\ \bibnamefont{Verma}},\ }%
  \bibfield{title}{%
  \enquote{\bibinfo {title} {{Statistical theory of magnetohydrodynamic
  turbulence: recent results}},}\ }%
  \bibfield{journal}{%
  \bibinfo {journal} {Phys. Rep.}\ }%
  \textbf{\bibinfo {volume} {401}},\ \bibinfo {pages} {229--380} (\bibinfo
  {year} {2004})\BibitemShut{NoStop}%
\bibitem{Bartello:JFM1994}%
  \BibitemOpen
  \bibfield{author}{%
  \bibinfo {author} {\bibfnamefont{P.}~\bibnamefont{Bartello}}, \bibinfo
  {author} {\bibfnamefont{O.}~\bibnamefont{Metais}},\ and\ \bibinfo {author}
  {\bibfnamefont{M.}~\bibnamefont{Lesieur}},\ }%
  \bibfield{title}{%
  \enquote{\bibinfo {title} {{Coherent structures in rotating three-dimensional
  turbulence}},}\ }%
  \bibfield{journal}{%
  \bibinfo {journal} {J. Fluid Mech.}\ }%
  \textbf{\bibinfo {volume} {273}},\ \bibinfo {pages} {1--29} (\bibinfo {year}
  {1994})\BibitemShut{NoStop}%
\bibitem{Godeferd:JFM1999}%
  \BibitemOpen
  \bibfield{author}{%
  \bibinfo {author} {\bibfnamefont{F.~S.}\ \bibnamefont{Godeferd}}\ and\
  \bibinfo {author} {\bibfnamefont{L.}~\bibnamefont{Lollini}},\ }%
  \bibfield{title}{%
  \enquote{\bibinfo {title} {{Direct numerical simulations of turbulence with
  confinement and rotation}},}\ }%
  \bibfield{journal}{%
  \bibinfo {journal} {J. Fluid Mech.}\ }%
  \textbf{\bibinfo {volume} {393}},\ \bibinfo {pages} {257--308} (\bibinfo
  {year} {1999})\BibitemShut{NoStop}%
\bibitem{Staplehurst:JFM2008}%
  \BibitemOpen
  \bibfield{author}{%
  \bibinfo {author} {\bibfnamefont{P.~J.}\ \bibnamefont{Staplehurst}}, \bibinfo
  {author} {\bibfnamefont{P.~A.}\ \bibnamefont{Davidson}},\ and\ \bibinfo
  {author} {\bibfnamefont{S.~B.}\ \bibnamefont{Dalziel}},\ }%
  \bibfield{title}{%
  \enquote{\bibinfo {title} {{Structure formation in homogeneous freely
  decaying rotating turbulence}},}\ }%
  \bibfield{journal}{%
  \bibinfo {journal} {J. Fluid Mech.}\ }%
  \textbf{\bibinfo {volume} {598}},\ \bibinfo {pages} {81--105} (\bibinfo
  {year} {2008})\BibitemShut{NoStop}%
\bibitem{Moisy:JFM2011}%
  \BibitemOpen
  \bibfield{author}{%
  \bibinfo {author} {\bibfnamefont{F.}~\bibnamefont{Moisy}}, \bibinfo {author}
  {\bibfnamefont{C.}~\bibnamefont{Morize}}, \bibinfo {author}
  {\bibfnamefont{M.}~\bibnamefont{Rabaud}},\ and\ \bibinfo {author}
  {\bibfnamefont{J.}~\bibnamefont{Sommeria}},\ }%
  \bibfield{title}{%
  \enquote{\bibinfo {title} {{Decay laws, anisotropy and
  cyclone{\textendash}anticyclone asymmetry in decaying rotating
  turbulence}},}\ }%
  \bibfield{journal}{%
  \bibinfo {journal} {J. Fluid Mech.}\ }%
  \textbf{\bibinfo {volume} {666}},\ \bibinfo {pages} {5--35} (\bibinfo {year}
  {2011})\BibitemShut{NoStop}%
\bibitem{Gallet:PF2014}%
  \BibitemOpen
  \bibfield{author}{%
  \bibinfo {author} {\bibfnamefont{B.}~\bibnamefont{Gallet}}, \bibinfo {author}
  {\bibfnamefont{A.}~\bibnamefont{Campagne}}, \bibinfo {author}
  {\bibfnamefont{P.~P.}\ \bibnamefont{Cortet}},\ and\ \bibinfo {author}
  {\bibfnamefont{F.}~\bibnamefont{Moisy}},\ }%
  \bibfield{title}{%
  \enquote{\bibinfo {title} {{Scale-dependent cyclone-anticyclone asymmetry in
  a forced rotating turbulence experiment}},}\ }%
  \bibfield{journal}{%
  \bibinfo {journal} {Phys. Fluids}\ }%
  \textbf{\bibinfo {volume} {26}},\ \bibinfo {pages} {035108} (\bibinfo {year}
  {2014})\BibitemShut{NoStop}%
\bibitem{Ranjan:JFM2014}%
  \BibitemOpen
  \bibfield{author}{%
  \bibinfo {author} {\bibfnamefont{A.}~\bibnamefont{Ranjan}}\ and\ \bibinfo
  {author} {\bibfnamefont{P.~A.}\ \bibnamefont{Davidson}},\ }%
  \bibfield{title}{%
  \enquote{\bibinfo {title} {{Evolution of a turbulent cloud under
  rotation}},}\ }%
  \bibfield{journal}{%
  \bibinfo {journal} {J. Fluid Mech.}\ }%
  \textbf{\bibinfo {volume} {756}},\ \bibinfo {pages} {488--509} (\bibinfo
  {year} {2014})\BibitemShut{NoStop}%
\bibitem{Verma:arxiv2017}%
  \BibitemOpen
  \bibfield{author}{%
  \bibinfo {author} {\bibfnamefont{M.~K.}\ \bibnamefont{Verma}}, \bibinfo
  {author} {\bibfnamefont{A.}~\bibnamefont{Kumar}}, \bibinfo {author}
  {\bibfnamefont{P.}~\bibnamefont{Kumar}}, \bibinfo {author}
  {\bibfnamefont{S.}~\bibnamefont{Barman}}, \bibinfo {author}
  {\bibfnamefont{A.~G.}\ \bibnamefont{Chatterjee}},\ and\ \bibinfo {author}
  {\bibfnamefont{R.}~\bibnamefont{Samtaney}},\ }%
  \bibfield{title}{%
  \enquote{\bibinfo {title} {{Energy fluxes and spectra for turbulent and
  laminar flows}},}\ }%
  \bibfield{journal}{%
  \bibinfo {journal} {Accepted in Fluid Dynamics}}%
   (\bibinfo {year} {2018})\BibitemShut{NoStop}%
\bibitem{Borue:PRL1993}%
  \BibitemOpen
  \bibfield{author}{%
  \bibinfo {author} {\bibfnamefont{V.}~\bibnamefont{Borue}},\ }%
  \bibfield{title}{%
  \enquote{\bibinfo {title} {{Spectral exponents of enstrophy cascade in
  stationary two-dimensional homogeneous turbulence}},}\ }%
  \bibfield{journal}{%
  \bibinfo {journal} {Phys. Rev. Lett.}\ }%
  \textbf{\bibinfo {volume} {71}},\ \bibinfo {pages} {3967} (\bibinfo {year}
  {1993})\BibitemShut{NoStop}%
\bibitem{Smith:PRL1993}%
  \BibitemOpen
  \bibfield{author}{%
  \bibinfo {author} {\bibfnamefont{L.M.}\ \bibnamefont{Smith}}\ and\ \bibinfo
  {author} {\bibfnamefont{V.}~\bibnamefont{Yakhot}},\ }%
  \bibfield{title}{%
  \enquote{\bibinfo {title} {{Bose condensation and small-scale structure
  generation in a random force driven 2D turbulence.}}.}\ }%
  \bibfield{journal}{%
  \bibinfo {journal} {Phys. Rev. Lett.}\ }%
  \textbf{\bibinfo {volume} {71}},\ \bibinfo {pages} {352--355} (\bibinfo
  {year} {1993})\BibitemShut{NoStop}%
\bibitem{Boffetta:EPL2007}%
  \BibitemOpen
  \bibfield{author}{%
  \bibinfo {author} {\bibfnamefont{G.}~\bibnamefont{Boffetta}}, \bibinfo
  {author} {\bibfnamefont{A.}~\bibnamefont{Cenedese}}, \bibinfo {author}
  {\bibfnamefont{S.}~\bibnamefont{Espa}},\ and\ \bibinfo {author}
  {\bibfnamefont{S.}~\bibnamefont{Musacchio}},\ }%
  \bibfield{title}{%
  \enquote{\bibinfo {title} {{Effects of friction on 2D turbulence: An
  experimental study of the direct cascade}},}\ }%
  \bibfield{journal}{%
  \bibinfo {journal} {EPL}\ }%
  \textbf{\bibinfo {volume} {71}},\ \bibinfo {pages} {590--596} (\bibinfo
  {year} {2005})\BibitemShut{NoStop}%
\bibitem{Rutgers:PRL1998}%
  \BibitemOpen
  \bibfield{author}{%
  \bibinfo {author} {\bibfnamefont{M.~A.}\ \bibnamefont{Rutgers}},\ }%
  \bibfield{title}{%
  \enquote{\bibinfo {title} {{Forced 2D turbulence: Experimental evidence of
  simultaneous inverse energy and forward enstrophy cascades}},}\ }%
  \bibfield{journal}{%
  \bibinfo {journal} {Phys. Rev. Lett.}\ }%
  \textbf{\bibinfo {volume} {81}},\ \bibinfo {pages} {2244} (\bibinfo {year}
  {1998})\BibitemShut{NoStop}%
\bibitem{Gotoh:PRE1998}%
  \BibitemOpen
  \bibfield{author}{%
  \bibinfo {author} {\bibfnamefont{T.}~\bibnamefont{Gotoh}},\ }%
  \bibfield{title}{%
  \enquote{\bibinfo {title} {Energy spectrum in the inertial and dissipation
  ranges of two-dimensional steady turbulence},}\ }%
  \bibfield{journal}{%
  \bibinfo {journal} {Phys. Rev. E}\ }%
  \textbf{\bibinfo {volume} {57}},\ \bibinfo {pages} {2984--2991} (\bibinfo
  {year} {1998})\BibitemShut{NoStop}%
\bibitem{Schorghofer:PRE2000}%
  \BibitemOpen
  \bibfield{author}{%
  \bibinfo {author} {\bibfnamefont{N.}~\bibnamefont{Schorghofer}},\ }%
  \bibfield{title}{%
  \enquote{\bibinfo {title} {Energy spectra of steady two-dimensional turbulent
  flows},}\ }%
  \bibfield{journal}{%
  \bibinfo {journal} {Phys. Rev. E}\ }%
  \textbf{\bibinfo {volume} {61}},\ \bibinfo {pages} {6572--6577} (\bibinfo
  {year} {2000})\BibitemShut{NoStop}%
\bibitem{Lindborg:PF2000}%
  \BibitemOpen
  \bibfield{author}{%
  \bibinfo {author} {\bibfnamefont{E.}~\bibnamefont{Lindborg}}\ and\ \bibinfo
  {author} {\bibfnamefont{K.}~\bibnamefont{Alvelius}},\ }%
  \bibfield{title}{%
  \enquote{\bibinfo {title} {The kinetic energy spectrum of the two-dimensional
  enstrophy turbulence cascade},}\ }%
  \bibfield{journal}{%
  \bibinfo {journal} {Phys. Fluids}\ }%
  \textbf{\bibinfo {volume} {12}},\ \bibinfo {pages} {945--947} (\bibinfo
  {year} {2000})\BibitemShut{NoStop}%
\bibitem{Ishihara:PF2001}%
  \BibitemOpen
  \bibfield{author}{%
  \bibinfo {author} {\bibfnamefont{T.}~\bibnamefont{Ishihara}}\ and\ \bibinfo
  {author} {\bibfnamefont{Y.}~\bibnamefont{Kaneda}},\ }%
  \bibfield{title}{%
  \enquote{\bibinfo {title} {Energy spectrum in the enstrophy transfer range of
  two-dimensional forced turbulence},}\ }%
  \bibfield{journal}{%
  \bibinfo {journal} {Phys. Fluids}\ }%
  \textbf{\bibinfo {volume} {13}},\ \bibinfo {pages} {544--547} (\bibinfo
  {year} {2001})\BibitemShut{NoStop}%
\bibitem{Chen:PRL2003}%
  \BibitemOpen
  \bibfield{author}{%
  \bibinfo {author} {\bibfnamefont{Q.}~\bibnamefont{Chen}}, \bibinfo {author}
  {\bibfnamefont{S.}~\bibnamefont{Chen}}, \bibinfo {author}
  {\bibfnamefont{G.~L.}\ \bibnamefont{Eyink}},\ and\ \bibinfo {author}
  {\bibfnamefont{K.~R.}\ \bibnamefont{Sreenivasan}},\ }%
  \bibfield{title}{%
  \enquote{\bibinfo {title} {{Kolmogorov's third hypothesis and turbulent sign
  statistics}},}\ }%
  \bibfield{journal}{%
  \bibinfo {journal} {Phys. Rev. Lett.}\ }%
  \textbf{\bibinfo {volume} {90}},\ \bibinfo {pages} {254501} (\bibinfo {year}
  {2003})\BibitemShut{NoStop}%
\bibitem{Borue:PRL1994}%
  \BibitemOpen
  \bibfield{author}{%
  \bibinfo {author} {\bibfnamefont{V.}~\bibnamefont{Borue}},\ }%
  \bibfield{title}{%
  \enquote{\bibinfo {title} {Inverse energy cascade in stationary
  two-dimensional homogeneous turbulence},}\ }%
  \bibfield{journal}{%
  \bibinfo {journal} {Phys. Rev. Lett.}\ }%
  \textbf{\bibinfo {volume} {72}},\ \bibinfo {pages} {1475--1478} (\bibinfo
  {year} {1994})\BibitemShut{NoStop}%
\bibitem{Tran:PRE2004}%
  \BibitemOpen
  \bibfield{author}{%
  \bibinfo {author} {\bibfnamefont{C.~V.}\ \bibnamefont{Tran}}\ and\ \bibinfo
  {author} {\bibfnamefont{J.~C.}\ \bibnamefont{Bowman}},\ }%
  \bibfield{title}{%
  \enquote{\bibinfo {title} {{Robustness of the inverse cascade in
  two-dimensional turbulence}},}\ }%
  \bibfield{journal}{%
  \bibinfo {journal} {Phys. Rev. E}\ }%
  \textbf{\bibinfo {volume} {69}},\ \bibinfo {pages} {036303} (\bibinfo {year}
  {2004})\BibitemShut{NoStop}%
\bibitem{chertkov:PRL2007}%
  \BibitemOpen
  \bibfield{author}{%
  \bibinfo {author} {\bibfnamefont{M.}~\bibnamefont{Chertkov}}, \bibinfo
  {author} {\bibfnamefont{C.}~\bibnamefont{Connaughton}}, \bibinfo {author}
  {\bibfnamefont{I.}~\bibnamefont{Kolokolov}},\ and\ \bibinfo {author}
  {\bibfnamefont{V.}~\bibnamefont{Lebedev}},\ }%
  \bibfield{title}{%
  \enquote{\bibinfo {title} {{Dynamics of energy condensation in
  two-dimensional turbulence}},}\ }%
  \bibfield{journal}{%
  \bibinfo {journal} {Phys. Rev. Lett.}\ }%
  \textbf{\bibinfo {volume} {99}},\ \bibinfo {pages} {084501} (\bibinfo {year}
  {2007})\BibitemShut{NoStop}%
\bibitem{Fischer:PF2009}%
  \BibitemOpen
  \bibfield{author}{%
  \bibinfo {author} {\bibfnamefont{P.}~\bibnamefont{Fischer}}\ and\ \bibinfo
  {author} {\bibfnamefont{C.-H.}\ \bibnamefont{Bruneau}},\ }%
  \bibfield{title}{%
  \enquote{\bibinfo {title} {{Wavelet-based analysis of enstrophy transfers in
  two-dimensional turbulence}},}\ }%
  \bibfield{journal}{%
  \bibinfo {journal} {Phys. Fluids}\ }%
  \textbf{\bibinfo {volume} {21}},\ \bibinfo {pages} {065109} (\bibinfo {year}
  {2009})\BibitemShut{NoStop}%
\bibitem{Ohkitani:PFA1991}%
  \BibitemOpen
  \bibfield{author}{%
  \bibinfo {author} {\bibfnamefont{K.}~\bibnamefont{Ohkitani}},\ }%
  \bibfield{title}{%
  \enquote{\bibinfo {title} {{Wave number space dynamics of enstrophy cascade
  in a forced two-dimensional turbulence}},}\ }%
  \bibfield{journal}{%
  \bibinfo {journal} {Phys. Fluids A}\ }%
  \textbf{\bibinfo {volume} {3}},\ \bibinfo {pages} {1598} (\bibinfo {year}
  {1991})\BibitemShut{NoStop}%
\bibitem{Babiano:JFM2007}%
  \BibitemOpen
  \bibfield{author}{%
  \bibinfo {author} {\bibfnamefont{A.}~\bibnamefont{Babiano}}\ and\ \bibinfo
  {author} {\bibfnamefont{Antonello}\ \bibnamefont{P.}},\ }%
  \bibfield{title}{%
  \enquote{\bibinfo {title} {{Coherent vorticies and tracer cascades in
  two-dimensional turbulence}},}\ }%
  \bibfield{journal}{%
  \bibinfo {journal} {J. Fluid Mech.}\ }%
  \textbf{\bibinfo {volume} {574}},\ \bibinfo {pages} {429--448} (\bibinfo
  {year} {2007})\BibitemShut{NoStop}%
\end{thebibliography}%

\end{document}